\documentclass[12pt]{article}
\pdfoutput=1
\usepackage[DIV13]{typearea}
\usepackage{graphicx}
\usepackage{amsmath}
\usepackage{amsfonts}
\usepackage{mathrsfs}
\usepackage{amssymb}
\usepackage{subfigure}
\usepackage{multicol}
\usepackage[usenames,dvipsnames]{color}
\usepackage{feynmp}
\usepackage{multirow}
\usepackage{array}
\usepackage{slashed}
\usepackage{units}
\usepackage{bbm}
\usepackage[normalem]{ulem}
\usepackage{framed}
\usepackage[left=.9in, right=.9in]{geometry}
\usepackage[titletoc,title]{appendix}
\newcolumntype{L}[1]{>{\raggedright\let\newline\\\arraybackslash\hspace{0pt}}m{#1}}
\newcolumntype{C}[1]{>{\centering\let\newline\\\arraybackslash\hspace{0pt}}m{#1}}
\newcolumntype{R}[1]{>{\raggedleft\let\newline\\\arraybackslash\hspace{0pt}}m{#1}}

\usepackage{hyperref}
\hypersetup{
 colorlinks,
 citecolor=Red,
 linkcolor=Blue,
 urlcolor=Blue}

\renewcommand{\L}{\mathcal{L}}
\renewcommand{\O}{\mathcal{O}}

\newcommand{\MeV}{{\rm MeV}}
\newcommand{\GeV}{{\rm GeV}}
\newcommand{\TeV}{{\rm TeV}}

\newcommand{\vev}[1]{\langle #1 \rangle}

\begin{document}

\vspace*{-0.2in}
\begin{flushright}
OSU-HEP-17-02~~~SLAC-PUB-16542~~~IFT-UAM-CSIC-16-051\\
FTUAM-16-21~~~FERMILAB-PUB-17-005-T~~~NSF-KITP-17-060
\end{flushright}

\vspace{0.5cm}

\begin{center}
{\Large\bf Flavor Gauge Models Below the Fermi Scale}\\
\end{center}

\vspace{0.5cm}
\begin{center}
{\large
{}~K.S. Babu$^{a,}$\footnote{E-mail: babu@okstate.edu},{}~
A. Friedland$^{b,}$\footnote{E-mail: alexfr@slac.stanford.edu; ORCID: http://orcid.org/0000-0002-5047-4680},{}~
{}~P.A.N. Machado$^{c,d,}$\footnote{E-mail: pmachado@fnal.gov; ORCID: http://orcid.org/0000-0002-9118-7354}, and
{}~I. Mocioiu$^{e,}$\footnote{E-mail: ium4@psu.edu}
}
\vspace{0.5cm}

\centerline{$^{a}${\it\small Department of Physics, Oklahoma State University, Stillwater, OK 74078, USA }}
\centerline{$^{b}${\it\small SLAC National Accelerator Laboratory, Stanford University, Menlo Park, CA, 94025, USA }}
\centerline{$^{c}${\it\small Departamento de F\'isica Te\'orica and Instituto de F\'{\i}sica Te\'orica, IFT-UAM/CSIC,}}
\centerline{{\it\small Universidad Aut\'onoma de Madrid, Cantoblanco, 28049, Madrid, Spain}}
\centerline{$^{d}${\it\small Theoretical Physics Department, Fermi National Accelerator Laboratory, Batavia, IL, 60510, USA}}
\centerline{$^{e}${\it\small Department of Physics, The Pennsylvania State University, University Park, PA 16802, USA }}
\end{center}
\vspace{0.6cm}

\begin{abstract}

The mass and weak interaction eigenstates for the quarks of the third generation are very well aligned, an empirical fact for which the Standard Model offers no explanation. We explore the possibility that this alignment is due to an additional  gauge symmetry in the third generation. Specifically, we construct and analyze an explicit, renormalizable model with a gauge boson, $X$, corresponding to the $B-L$ symmetry of the third family. Having a relatively light (in the MeV to multi-GeV range), flavor-nonuniversal gauge boson results in a variety of constraints from different sources. By systematically analyzing 20 different constraints, we identify the most sensitive probes: kaon, $B^+$, $D^+$ and Upsilon decays, $D-\bar{D}^0$ mixing, atomic parity violation, and neutrino scattering and oscillations. For the new gauge coupling $g_X$ in the range $(10^{-2} - 10^{-4})$ the model is shown to be consistent with the data. Possible ways of testing the model in $b$ physics, top and $Z$ decays, direct collider production and neutrino oscillation experiments, where one can observe nonstandard matter effects, are outlined. The choice of leptons to carry the new force is ambiguous, resulting in additional phenomenological implications, such as  non-universality in semileptonic bottom decays. The proposed framework provides interesting connections between neutrino oscillations, flavor and collider physics.
\end{abstract}

\newpage

\section{Introduction}

One of the long-standing puzzles of the Standard Model (SM) is the origin of flavor: understanding why all fermion fields come in three families, or generations. Within each family the gauge quantum numbers are perfectly coordinated to cancel all 10 potential gauge anomalies (see for {\it e.g.}, \cite{Peskin:1995ev}), ensuring the theoretical consistency of the SM as a chiral gauge theory~\cite{Gross:1972pv}. In contrast, the SM has no similar consistency condition that would require combining particles of \emph{different} generations.
In this sense, while every member of a given family is  indispensable for making that family consistent, the different families do not seem to have a need for one another. 

In searching for answers to the fundamental questions of flavor physics, the first step is to understand the physical properties of the generations. Here again Nature offers a puzzle: in the SM the families are identical copies of each other in some characteristics, but not all. Specifically, partners from different generations are thought to have exactly the same (\emph{universal}) gauge interactions, while their Yukawa couplings to the Higgs field are vastly different, as reflected by their masses. Perhaps the Yukawa and gauge interactions are unrelated? Yet, the pattern of the mixing angles in the CKM matrix does not appear random. This is especially so for the third family quarks, which are the most massive of the six and mix little with the first two generations. Explicitly, the top quark (a mass eigenstate) upon emitting the $W$ gauge boson becomes very nearly the bottom quark mass eigenstate. This accurate alignment of the flavor and mass bases seems like an odd coincidence and suggests some underlying connection between the gauge and Yukawa interactions.

Here, we explore a possibility that this alignment of the eigenstates is a sign that the gauge interactions are actually not strictly universal. The idea is simple: if the third generation is charged under an additional gauge group, it cannot mix with the first two using the SM Higgs field. Notice that this is merely a statement of charge conservation, so that the new gauge coupling need not be large. This implies that once the new gauge group is broken somewhere in the vicinity of the weak scale, as we discuss below, the mediator of the new force can be quite light. This may sound dangerous from flavor violation constraints, but as we will see, there is a well-defined allowed region of the parameter space.

How do we choose the new gauge interaction to assign to the third generation? As our guiding principle, we wish to preserve the elegant feature of the SM outlined above: that all anomalies cancel \emph{within a generation}. It is well known that the simplest gauge group with such properties is based on the difference of the baryon and lepton numbers, $U(1)_{B-L}$, provided one adds a right-handed sterile neutrino to cancel the cubic anomaly. Thus, \emph{this paper is devoted to the phenomenology of the weakly gauged $U(1)_{B-L}^{(3)}$}.

Let us briefly review how our framework is different from the existing literature. The observation that $U(1)_{B-L}$ is anomaly free and can be gauged has been made four decades ago and has been studied in numerous contexts. The classical framework \cite{Pati:1974yy,Marshak:1979fm,Wilczek:1979et,Mohapatra:1980qe} considers this symmetry to be flavor-universal and broken at a high scale, so that the lepton-number violating (LNV) Majorana mass for the sterile neutrino is generated and LNV effects are then transmitted to the light neutrinos via the seesaw mechanism. More recently, $B-L$ was considered to be broken at the low scale, but again in a strictly flavor-universal setup~\cite{Nelson:2007yq}. Some additional constraints on this low-scale mediator were obtained in~\cite{Harnik:2012ni}. None of these cases consider flavor-nonuniversality. New light physics is also flavor-universal in another class of models, those involving a dark photon, which interacts with the SM via kinetic mixing~\cite{holdom}. Finally, there have been ideas to study flavor-dependent, horizontal gauge symmetries~\cite{Wilczek:1978xi} (for related discussions in a dynamical electroweak symmetry breaking context, see e.g. Ref.~\cite{Appelquist}). Gauging the symmetry based on $L_{\mu}-L_{\tau}$ \cite{He:1991qd} has attracted quite a bit of interest in recent years \cite{Baek:2001kca,Ma:2001md,Salvioni:2009jp,Heeck:2011wj,Harigaya:2013twa,Carone:2013uh,Altmannshofer:2014cfa,Farzan:2015doa,Farzan:2015hkd}. While such new interactions would be also anomaly-free, the cancellation is achieved between generations. This class of model is very different from ours, both in terms of its philosophy and its physics.

Let us outline some of the generic consequences of gauging $U(1)_{B-L}^{(3)}$. The most obvious one is the existence of an extended Higgs sector. Indeed, in addition to the Higgs field with the SM quantum numbers (henceforth $\phi_{2}$), a new field, $\phi_{1}$, charged under the new gauge symmetry is required, to allow for nonzero mixing between the third family and the first two. As we will see, to make the theory phenomenologically viable, one also needs to introduce another scalar field, $s$, that is a singlet under $SU(3)_{c} \times SU(2)_{L}\times U(1)_{Y}$, but is charged under $U(1)_{B-L}^{(3)}$. Together, the
vacuum expectation values of $\phi_{1}$ and $s$ will spontaneously break $U(1)_{B-L}^{(3)}$, giving a mass $M_{X}$ to the new gauge boson $X$. Moreover, the vacuum expectation value ({\it VEV}) of $\phi_1$ will mix the $X$ with the electroweak $Z$ boson.

Because of the $X-Z$ mixing, the new force will actually couple not only to the third generation, but also to the first two, with appropriate suppression factors. This is the second generic consequence of our framework. The model predicts additional neutral currents and one has to carefully ensure existing tight bounds are not violated. This means analyzing a plethora of constraints and identifying the dominant ones for different values of the mediator mass $M_{X}$. Needless to say, we are required to dispense with the effective field theory descriptions that are usually assumed when analyzing new flavor physics constraints (see for example \cite{D'Ambrosio:2002ex}). When the new gauge boson is light, one should of course keep it in the
low energy spectrum as a dynamical field, all the way down to energy
scales below its mass.

The analysis of the neutral currents also extends to the lepton sector. Here, we find our third general prediction: the neutrinos will interact non-universally with matter and the MSW potential will gain additional terms. Thus, our framework is a model of neutrino non-standard interactions (NSI), which have been of phenomenological interest to the oscillation community for a number of years~\cite{Wolfenstein:1977ue,Valle:1987gv,Roulet:1991sm,Guzzo:1991hi,GonzalezGarcia:2001mp,Fornengo:2001pm,Davidson:2003ha,Friedland:2004pp,Friedland:2005vy,Antusch:2008tz,Gavela:2008ra,GonzalezGarcia:2011my,Friedland:2011za,Friedland:2012tq,Gonzalez-Garcia:2015qrr}. It is remarkable that in some parts of the parameter space neutrino oscillations already provide important constraints on the model. It is also remarkable that these NSI effects probe a certain combination of the Higgs vacuum expectation values (VEVs) at the weak-scale and not the light mass $M_{X}$. Of course, the effects are communicated to our sector via $X$, but the value of $M_{X}$ drops out from the oscillation potential.

Another important class of constraints, in which the mass $M_{X}$ drops out, is made up of processes dominated by the longitudinal mode of $X$. As seen below, the relevant mode is actually properly understood as the Goldstone from the extended Higgs sector that is eaten by $X$. As a consequence, the relevant rates depend only on the Yukawa couplings and not on $g_{X}$. These bounds therefore apply even in the limit of infinitesimally gauged (global) $U(1)_{B-L}^{(3)}$.

It should be by now obvious that the analysis of this model is by necessity very rich: we investigate over twenty potential constraints. Of these, we identify a subset of essential bounds: they come from $\Upsilon$, Kaon and $B$ decays, $D$ decays and $D-\bar D$ oscillations, atomic parity violation, neutrino oscillations and electroweak precision observables. Each of these becomes dominant in some parts of the parameter space. Of course, to be sure that the other dozen constraints are subdominant, we are required to evaluate them as well. To keep the scope of the paper finite, we deliberately do not include any discussions of the astrophysical constraints here. We also do not consider in details certain model-building aspects and collider constraints. They will be covered in separate  publications.

Before turning to our main presentation, two important comments about the lepton sector of the theory have to be made. First, what we call ``the third generation leptons'' is strictly speaking {\it a priori} ambiguous: since there are no gauge bosons connecting the third generation quarks with the lepton sector the same way the top and bottom quarks are connected, we do not know that it is the $\tau$ lepton that has to be assigned $U(1)_{B-L}^{(3)}$. In fact, any linear combination of the leptons from the three generation can be used to cancel the anomalies of the third family quarks and hence any such combination could be made charged under the new gauge group. We stick with $\tau$ and $\nu_\tau$ as the ``the third generation leptons'' only for definiteness. This choice is  made to once again keep the scope of the present paper manageable.

Second, so far we have avoided any mention of the leptonic mixing, which is clearly different from the pattern in the quark sector. This qualitative difference already points to the different physical origin of the neutrino masses compared to those of quarks. Indeed, we will see this when we briefly discuss the framework for neutrino masses below. Our masses are of Majorana type and can be obtained from seesaw-type relations. Notice that one important difference compared to the classical seesaw is that the right-handed $B-L$ partner neutrino (required by anomaly cancellation) lives near the scale of the Higgs VEVs, where the gauge symmetry is broken. Therefore, there are physical arguments to expect the neutrino mass mechanism in our model to be potentially within reach of collider physics.

The rest of the paper is organized as follows.  In Sec.~\ref{sec:model} we present and analyze the
$U(1)_{B-L}^{(3)}$ model. Sec.~\ref{sec:pheno} provides a summary of the main experimental constraints
on the model. Sec.~\ref{sec:appendix} discusses other low energy constraints on the model.  In Sec.~\ref{sect:conclusions} we discuss some important overall consequences of our findings and provide an  outlook for future searches for this scenario.

\section{The \boldmath{$U(1)_{B-L}^{(3)}$} model}
\label{sec:model}

The model we study is based on the Standard Model symmetry extended by a
$U(1)_{B-L}^{(3)}$ gauge symmetry.  $B-L$ symmetry is anomaly free
for each generation of fermions, provided that a right-handed neutrino
is introduced.  Thus the $U(1)_{B-L}^{(3)}$ charges of fermions in our
extended model are $(Q_{3L}, u_{3R}, d_{3R}): 1/3$, $(\ell_{3L}, e_{3R}, \nu_{3R}): -1$,
with all fermions of the first two families carrying zero charges.  This is
true for $(\nu_{1R}, \nu_{2R})$ as well, and as a result these states could in principle acquire
large Majorana masses and decouple from the low energy theory.  We do
use these states for neutrino mass generation through effective seesaw operators.

The gauge boson associated with $U(1)_{B-L}^{(3)}$ is denoted $X$, and we shall
be interested in the case where $M_X$ is in the MeV--multi-GeV range.
Flavor effects have been
widely studied when $M_X$ is larger than the electroweak scale, while below about 100 keV stellar cooling bounds
typically require the gauge coupling to be so small that such an $X$ boson would be of
little interest for flavor phenomenology. Although the mass of $X$ is in the
MeV--multi-GeV range, the scale of $U(1)_{B-L}^{(3)}$ symmetry breaking could
be several hundred GeV, which is what we shall take as our benchmark value. This is possible owing
to the smallness of the gauge coupling $g_X$.

A minimal scalar sector for the model consists of two Higgs doublets, $\phi_2$ with zero $U(1)_{B-L}^{(3)}$ charge
and $\phi_1$ carrying $U(1)$ charge of $1/3$,
as well as a SM singlet field $s$.  The $U(1)_{B-L}^{(3)}$ charges of the scalars  are listed in Table
\ref{tab:charge}.  $\phi_2$ is the Higgs doublet that generates diagonal mass terms
for the quarks and leptons, while $\phi_1$ induces off-diagonal quark mixing terms involving the third family.
The field $s$ is needed for consistent phenomenology as well as for inducing
neutrino mixings via simple effective operators. As we shall see, without the singlet field, the
contributions to non-standard neutrino oscillations from the $X$ gauge boson will exclude the model.
The $U(1)_{B-L}^{(3)}$ charge of $s$ field is uniquely fixed to be $1/3$ or $1/6$,
other choices would lead to an enhanced global $U(1)$ symmetry in the Higgs potential, resulting
in an unwanted pseudo-Goldstone boson. (A term of the type $\phi_1^\dagger\phi_2 s$ or $\phi_1^\dagger\phi_2 s^2$ would break such a global symmetry explicitly and give mass to the Goldstone boson.)  We shall focus on $s$ charge being $1/3$, which leads to a slightly
simpler neutrino mass generation scheme.

Since the Higgs doublet $\phi_1$ carries both $U(1)_Y$ and $U(1)_{B-L}^{(3)}$
charges, when its neutral component acquires a vacuum expectation value it will induce mixing between
the $Z$ and the new gauge boson $X$. As the new symmetry is an Abelian $U(1)$,
the model also admits the possibility of kinetic mixing between the hypercharge
gauge boson and the $X$ boson \cite{holdom}.

\begin{table}[h]
  \begin{center}
    \begin{tabular}{|c|c|c|c|} \hline
  & $\phi_1$ & $\phi_2$ & $s$\\ \hline
$SU(2)_L$& \bf 2 & \bf 2 & \bf 1 \\ \hline
$U(1)_Y$& +1 & +1 & 0 \\ \hline
$U(1)_{B-L}^{(3)}$ & +1/3 & 0 & +1/3 \\ \hline
    \end{tabular}
  \end{center}	
  \caption{Scalar fields and their charges under the Standard
    Model gauge group and the $U(1)_{B-L}^{(3)}$ gauge symmetry. In our notation,
    the $U(1)_{B-L}^{(3)}$ charge of the third family quarks is $+1/3$, while that for the third
    family leptons is $-1$. The first two families of fermions have zero $U(1)_{B-L}^{(3)}$ charges.}
\label{tab:charge}
\end{table}

\subsection{The Yukawa sector}

Since the third family quarks carry a nonzero $U(1)_{B-L}^{(3)}$ charge while the first two families
do not, the Yukawa couplings that would induce three family quark mixing should
involve both doublets $\phi_1$ and $\phi_2$.  The $\phi_1$ field is introduced for the purpose of
inducing quark mixing with the third family.
The Yukawa Lagrangian for the quarks is given by
\begin{equation}
\label{eq:interaction}
\L^q_{yuk} = \overline{\mathbf{Q}}_L
  \left(\begin{array}{ccc}
    y_{11}^u \widetilde\phi_2 & y_{12}^u \widetilde\phi_2 & y_{13}^u \widetilde\phi_1\\
    y_{21}^u \widetilde\phi_2 & y_{22}^u \widetilde\phi_2 & y_{23}^u \widetilde\phi_1\\
    0                        & 0                        & y_{33}^u \widetilde\phi_2\\
  \end{array}\right)\mathbf{u}_R
+ \overline{\mathbf{Q}}_L
  \left(\begin{array}{ccc}
    y_{11}^d \phi_2 & y_{12}^d \phi_2 & 0\\
    y_{21}^d \phi_2 & y_{22}^d \phi_2 & 0\\
    y_{31}^d \phi_1 & y_{32}^d \phi_1 & y_{33}^d \phi_2\\
  \end{array}\right)\mathbf{d}_R + \text{h.c.}
\end{equation}
Here the bold symbols stand for vectors in generation
space,  and $\widetilde\phi_i\equiv i \sigma_2 \phi_i^*$ with $\sigma_2$ being
the second Pauli matrix.
The simultaneous presence of $\phi_1$ and $\phi_2$ in the Yukawa couplings of the
up-quarks (and similarly for the down-quarks) would imply that there are Higgs-mediated
FCNC processes in the model. We shall see that
these processes are within acceptable limits, provided that the neutral Higgs bosons have
masses of order hundred GeV.

As only the third family carries the new $U(1)_{B-L}^{(3)}$ charge, the
Cabibbo angle can be generated without inducing any FCNC
mediated by neutral scalar bosons or the $X$ gauge boson. We thus make 1-2 rotations in
both the up- and down- quark sectors, thereby inducing a nonzero $(1,2)$ entry in the CKM matrix.
The other CKM matrix elements $V_{ub}$ and $V_{cb}$ can be generated from the rotated mass matrices
which can be written in the form
\begin{equation}\label{eq:mixingmatrix}
R_{12}^{ uL}.M_u.R_{12}^{ uR\dagger} =  \left(\begin{array}{ccc}
    m_u^0 & 0 & V_{ub}^0 m_t^0\\
    0 & m_c^0 & V_{cb}^0 m_t^0\\
    0 & 0 & m_t^0
\end{array}\right)
\quad  \text{and} \quad
R_{12}^{ dL}.M_d.R_{12}^{ dR\dagger} =   \left(\begin{array}{ccc}
    m_d^0 & 0 & 0\\
    0 & m_s^0 & 0\\
    a m_b^0 & b m_b^0 & m_b^0
\end{array}\right)
\end{equation}
where $R_{ij}$  parametrizes an $i-j$ rotation in terms  of a mixing angle and a
phase.  While these forms are quite general, we shall approximate
$m_i^0$ in Eq. (\ref{eq:mixingmatrix}) to be nearly equal to the physical eigenvalue
$m_i$ and $V_{ij}^0$ to be nearly equal to the actual CKM mixing element $V_{ij}$.

The down quark mass matrix given in Eq. (\ref{eq:mixingmatrix}) is diagonalized by right-handed
rotations alone, with the left-handed mixing matrix being very close to an identity matrix.
Thus $V_{cb}$ and $V_{ub}$ should arise primarily from the up-quark sector.  The FCNC constraints
arising from the down-quark sector are more severe compared to those arising from the up-quark
sector.  Assuming that $m_b^0\simeq
m_b$, $B_d-\bar{B}_d$ mixing mediated by the neutral scalar bosons
sets a limit $a\lesssim 3\times10^{-3}/\tan\beta$ for
scalar masses of order 100~GeV, while $B_s-\bar{B}_s$ mixing constrains
$b\lesssim 10^{-2}/\tan\beta$  on the parameters $a$ and $b$ appearing in the down quark mass matrix
in Eq. (\ref{eq:mixingmatrix}) (see Sec.~\ref{sec:appendix} for
details). Here we have defined $\tan\beta \equiv v_2/v_1$.  Similar constraints are obtained from the decays
 $B_d\to X\gamma\to e^+e^-\gamma$~\cite{Agashe:2014kda} and $B_s\to X\to\mu^+\mu^-$.
  More importantly, off-diagonal couplings $Xdb$ and $Xsb$ would contribute to the total width of $B_d$ and $B_s$, as well as to $B^+\to\pi^+e^+e^-$ and $B^+\to\pi^+\mu^+\mu^-$. The first and second widths would constrain $g_X (b/V_{cb}) < 2.8\times 10^{-6}(M_X/100~\MeV)$ and $g_X (a/V_{ub}) < 2.9\times 10^{-5}(M_X/100~\MeV)$, while the last processes would lead to
\begin{equation}
  g_X\frac{a}{V_{ub}}<1.8\times 10^{-10}\frac{M_X/100~\MeV}{\sqrt{BR(X\to e^+e^-)}},
  \quad g_X\frac{a}{V_{ub}}<3.8\times 10^{-10}\frac{M_X/100~\MeV}{\sqrt{BR(X\to\mu^+\mu^-)}},
\end{equation}
  see Appendix~\ref{Appendix} for details.

With these constraints, the parameters $a$ and $b$ in Eq. (\ref{eq:mixingmatrix})
cannot significantly contribute to the generation of CKM mixing angles $V_{cb}$ and
$V_{ub}$, which we shall thus ignore.  Notice that FCNCs will be induced in the down sector at loop level, and that is particularly important for Kaon decays, as we will see in Sec.~\ref{sec:pheno}.
Within these assumptions,
the left-handed rotations that diagonalize $M_u$ and $M_d$ are given by (in
a basis where the 1-2 up-sector is already diagonal, i.e., with $R_{12}^{uL}$, $R_{12}^{uR}$ being identity matrices)
\begin{align}
  &V_u^L = R_{23}^{ uL}(V_{cb})R_{13}^{ uL}(V_{ub}),\\
  &V_d^{L\dagger} = R_{12}^{ dL}(V_{us})^\dagger.
\end{align}
 If the $X$ charge of the scalars are instead chosen to be $-1/3$, the Yukawa Lagrangian for up-type quarks and down-type quarks (\ref{eq:interaction}) would be interchanged. That would suggest the generation of $V_{ub}$ and $V_{cb}$ in the down sector, which would lead to strong constraints in $g_X$, as discussed above. We do not pursue such possibility in this manuscript.
The quark mixing matrix is given by $V_{\rm CKM} = V_u^{ L}V_d^{
  L\dagger}$.  It can be readily checked that a CP violating phase of the correct magnitude is obtained from complex entries of the mass
  matrices.  It follows from
Eq.~(\ref{eq:mixingmatrix}) that any FCNC effects induced by scalar boson
exchanges would be weighted by  $V_{ub}$ and
$V_{cb}$ in the top sector where the experimental constraints are meager,
and by $V_{ub}V_{cb}$ in the $u-c$ sector. This suppression factor will be
sufficient to avoid the stringent $D^0-\overline {D^0}$ mixing bounds and mitigate the effect on $D^+$ decays with $\Delta C=1$, as we will see
in Sec.~\ref{sec:pheno}.

In the charged lepton sector Yukawa couplings between the third
and the first two families are strictly forbidden owing  to the charge
assignment and minimality of the Higgs sector of the model.
Charged lepton masses arise through the Yukawa
Lagrangian involving the  $\phi_2$ scalar only and is given by
\begin{equation}
{\cal L}_{yuk}^{\ell} =
  y^\ell_{ij}\overline{L}_i\phi_2 \ell_{Rj},
\end{equation}
with $y_{ij}=0$ for $ij=13,23,31,32$. We see that the leptonic mixing angle $\theta_{12}^\ell$ could
be generated from here, but not $\theta_{23}^\ell$ and $\theta_{13}^\ell$.
There are no FCNC processes mediated by the Higgs bosons, since the Yukawa coupling matrix
is proportional to the charged lepton mass matrix.  There are also no FCNC processes mediated by the $X$ gauge boson, since
the mass eigenbasis and the flavor eigenbasis coincide for the charged leptons.
The complete absence of tree-level
FCNC in the charged lepton sector is a compelling feature of the model,
protecting it from the severe bounds that could have arisen from flavor changing muon and tau
decays.

Neutrino mass generation calls for additional physics which can however reside at a higher scale. In
the minimal setup considered here, we can infer neutrino masses as arising
from effective operators via a generalized seesaw mechanism.
For the 1-2 sector of the effective Majorana
matrix of the light neutrinos, the usual dimension--5 operator can be built (with $\tilde L_i\equiv i \tau_2 L_i^*$):
\begin{equation}  \label{d=5}
  \frac{1}{\Lambda} \left(\bar L_{1,2}\tilde\phi_2\right)
  \left(\phi_2^\dagger \tilde L_{1,2}\right),
\end{equation}
while the mixing responsible for $\theta_{13}^\ell$ and $\theta_{23}^\ell$ should
come from a dimension--6 operator
\begin{equation}  \label{d=6}
  \qquad
  \frac{1}{\Lambda^2} \left(\bar L_{3} \tilde\phi_1\right)
  \left(\phi_1^\dagger \tilde L_{1,2}\right)s^*.
\end{equation}
These operators could be generated by exchanging singlet neutrinos
with $U(1)_{B-L}^{(3)}$ charges 0, $\pm1/3$ and $\pm2/3$. The first of
those can be identified as the usual right-handed neutrinos of the
first two families, while the remaining two are singlet fermions which
are vector-like under $U(1)_{B-L}^{(3)}$. Note
that the right-handed neutrino $\nu_{3R}$ with $U(1)_{B-L}^{(3)}$ charge $-1$ will mix with
the vector-like component with charge $\pm 2/3$ via the Yukawa coupling
$\nu_{3R} n_{2/3}s$ once the $s$ field acquires a VEV. Thus there are no light  sterile
neutrinos in the model, provided that the vector-like singlet neutrinos are not too heavy
(otherwise the mass of $\nu_{3R}$ will become small via a seesaw suppression factor).

Since all neutrino mixing angles are relatively large, the mass matrix elements coming from
the dimension--5 and the dimension--6 operators should be comparable. If the
singlet neutrinos that are integrated out have masses not far above
the TeV scale, so that they also do not introduce an additional hierarchy
problem for the Higgs boson mass \cite{vissani}, then these different contributions to
light neutrino masses would be of the same order.
Besides, as $\nu_{3R}$ is needed to cancel anomalies, its mass cannot be decoupled from the theory:
the mass of this state should be close to or below $v_s$. As we will see later, typical values for $v_s$
lie between 100-1000~GeV, assuming no  new hierarchy problem in the scalar sector is introduced in the model.
This provides a deeper reason for why at least part of the sterile neutrino spectrum should be
accessible at the LHC. The LHC phenomenology of the neutrino mass generation sector may be pursued
in a future manuscript.

\subsection{The gauge boson sector}

Now we turn our attention to the gauge boson sector.
We adopt the convention $q=I_3+Y/2$ for the hypercharge, where $q$ is the electric charge, $I_3=0,\pm1/2$ for $SU(2)_L$
singlet and doublet fields, and $Y$ is the hypercharge. The gauge kinetic terms for the scalar fields are given by
$\sum_i|D_\mu \phi_i|^2 + |D_\mu s|^2$ where the
covariant derivatives are defined as
\begin{equation}
  D_\mu\phi_i = \left(\partial_\mu -i g \frac{\tau_i}{2}W^i_\mu
  - i g^\prime \frac{Y}{2}B_\mu -i g_X q_X X^0_\mu \right)\phi_i,~~ D_\mu s = \partial_\mu s - i g_X q_X s.
\end{equation}
When the scalar fields acquire VEVs, $SU(2)_L\times U(1)_Y\times
U(1)_{B-L}^{(3)}$ symmetry breaks spontaneously down to $U(1)_{em}$.  Since the doublet
field $\phi_1$ is charged under both $Y$ and $U(1)_{B-L}^{(3)}$, its VEV will induce mixing between the $Z$ and the
new gauge boson $X$. In the absence of kinetic mixing the
gauge boson mass-squared matrix is given as
(in the basis ($Z^0, X^0$) where the $0$ subscript indicates a state before $Z-X$ mixing)
\begin{equation}\label{eq:mass-gauge}
  M^2_{\rm gauge} = \frac{1}{4}\left(\begin{array}{cc}
(g^2+g^{\prime 2})v^2   &   -2\sqrt{g^2+g^{\prime 2}}g_X v_1^2/3\\
-2\sqrt{g^2+g^{\prime 2}}g_X v_1^2/3 & 4g_X^2(v_1^2+v_s^2)/9
\end{array}\right).
\end{equation}
Here $v_1, v_2, v_s$ are the VEVs of $\phi_1$, $\phi_2$, and $s$,
respectively, with $v_1^2+v_2^2 \equiv v^2=(246~{\rm GeV})^2$.  The photon is
still the combination $A_\mu = c_w B_\mu + s_w W^3_\mu$ ($c_w = \cos\theta_w, s_w = \sin\theta_w$, $\tan\theta_w = g'/g$),
while the physical $Z$ and $X$ boson eigenstates are given by (ignoring terms of order $\O(g_X^2)$),
\begin{align}
  Z_\mu &\simeq -s_w B_\mu + c_w W^3_\mu -  s_X X^0_\mu,\\
  X_\mu &\simeq s_X(-s_w B_\mu + c_w W^3_\mu) +  X^0_\mu,
\end{align}
with the $Z-X$ mixing angle $s_X$ defined as
\begin{equation}
  s_X\equiv \frac{2}{3}\frac{g_X}{\sqrt{g^2+g^{\prime 2}}}\frac{v_1^2}{v^2}.
\end{equation}
We observe that it is the VEV of $\phi_1$ that induces the $Z-X$ mixing, and that
$s_X$ is proportional to $g_X$ and $v_1$. The mass of the $X$ gauge boson is obtained as
\begin{equation} \label{eq:mx}
  M_X^2 = \frac{1}{9}g_X^2\left( \frac{v_1^2v_2^2}{v^2} + v_s^2 \right).
\end{equation}
Notice that a nonzero $v_s$ can only raise
$M_X$. When $v_1$ and $v_2$ are comparable,
$M_X$ is essentially fixed in terms of $v_s$, while for large $\tan\beta$ there
is some dependence on $v_1$ and $v_2$ as well. Then, for a given $g_X$, Eq.~(\ref{eq:mx}) defines
a minimum mass for the $X$ boson.

As will be seen later, the longitudinal mode $X_L$ plays a prominent role on the phenomenology,
particularly in the case of light $X$ (with respect to the scale of the process in question). In such
case, the equivalence theorem implies that $X_L$ can be substituted by its corresponding
Goldstone boson $G_X$. It is easy to see that $G_X$ is given by
\begin{equation}
  G_X = \frac{1}{3}\frac{g_X}{M_X v^2}\left[-v_1 v_2^2 \,{\rm Im}(\phi_1^0) + v_1^2 v_2\, {\rm Im}(\phi_2^0) - v^2 v_s\, {\rm Im}(s^0)\right].
\end{equation}
Some of the Goldstone boson couplings will be particularly important, namely,
\begin{align}\label{eq:goldstone-fermions}
  \mathcal{L}_{G_X}=\,& iG_X\frac{g_X}{3}\frac{m_t}{M_X}\left[ - \frac{v_1^2}{v^2} \bar{t}\gamma_5 t
    +V_{cb}(\bar c_L t_R - \bar t_R c_L)
    +V_{ub}V_{cb}(\bar c_L u_R - \bar u_R c_L)\right]\nonumber\\
    &\hspace{7cm}-iG_X\frac{g_X}{3}\frac{m_\tau}{M_X}\frac{v_1^2}{v^2}\bar\tau\gamma_5\tau+\dots
\end{align}
We shall use these couplings when deriving the constraints from decays of various particles into longitudinal modes of $X$ boson.

The gauge boson kinetic terms allow for mixing between $X_{\mu\nu}$
and $B_{\mu\nu}$ parametrized by $\varepsilon$. These are given by
\begin{align}
  \L_{kin} &= -\frac{1}{4}W^3_{\mu\nu}W^{3\mu\nu}  -\frac{1}{4}B_{\mu\nu}B^{\mu\nu}
  -\frac{1}{4}X_{\mu\nu}X^{\mu\nu}  +\frac{\varepsilon}{2}X_{\mu\nu}B^{\mu\nu}\\
  &=-\frac{1}{4}A_{\mu\nu}A^{\mu\nu}  -\frac{1}{4}Z_{\mu\nu}Z^{\mu\nu}
  -\frac{1}{4}X_{\mu\nu}X^{\mu\nu}
  +\frac{\varepsilon}{2}X_{\mu\nu}(c_w A^{\mu\nu}-s_wZ^{\mu\nu}) + \O(\varepsilon^3).
\end{align}
To obtain canonical kinetic terms for the gauge bosons, up to
$\O(\varepsilon^3)$, the photon and the $X$ fields can be
redefined as~\cite{Baumgart:2009tn}
\begin{align}
  A_\mu &\to A_\mu + \varepsilon c_w X_\mu,\\
  X_\mu &\to X_\mu - \varepsilon s_w Z_\mu.
\end{align}
The effect of the photon field shift is only to couple the standard
electromagnetic current to $X$, with the coupling strength being  $\varepsilon c_w$. The
$X$ field shift has two effects. First, it couples the $X$ current to
the $Z$ charge, so the $Z$ couplings to particles that are charged under the
new symmetry are slightly modified. Second, as $X$ is massive, its
shift gives rise to a $Z-X$ mass term $-2\varepsilon s_w
M_X^2$. Assuming $M_X\ll M_Z$, a small rotation by $\varepsilon
M_X^2/M_Z^2$ is required to have diagonal mass terms for the $Z$
and $X$ bosons. Due to the additional suppression factor $M_X^2/M_Z^2$,
this rotation is not significant, and we shall
neglect this effect. It is important to notice that the non-unitary
character of the shift assures the absence of millicharged particles:
although electrically charged particles acquire small $X$ charges, the opposite, viz., particles
charged under $X$ acquiring small electric charge,  does not
happen.

Since the $U(1)_{B-L}^{(3)}$ gauge interaction distinguishes flavor,
it leads to FCNCs. In the flavor basis the $X$ interactions to SM
fermions are given by
\begin{equation}\label{eq:Xcoupling}
  \L_{ffX}=c_\alpha\bar f_\alpha \gamma_\mu f_\alpha X^\mu,\quad{\rm with}\quad
  c_\alpha = q_\alpha c_w e\,\varepsilon
  + \left(g_X q_\alpha^X + s_X\sqrt{g^2+g^{\prime 2}}q^Z_\alpha\right),
\end{equation}
where $q_\alpha$, $q_\alpha^X$, and
$q_\alpha^Z=I_3^\alpha-s_w^2q_\alpha$, are the electric charge, the
$X$ charge and the $Z$ charge, respectively, of the fermion
$\alpha$. Notice that, as $c_\alpha$ depends on the chirality of the
field, it is not possible to have an accidental cancellation between
$\varepsilon$ and $g_X$ for both $L$ and $R$ components of any
particle. The relative sign (and magnitude) between
$\varepsilon$ and $g_X$ is physically observable.

We can understand the FCNC processes induced by the $X$ gauge boson by writing the
non-universal piece of the interaction explicitly as
\begin{equation}
  \L_{\rm X-FCNC} =  \frac{g_X}{3} \overline{\mathbf{Q}}_L
  \left(\begin{array}{ccc}
    0 & 0 & 0\\
    0 & 0 & 0\\
    0 & 0 & 1
\end{array}\right)\gamma^\mu \mathbf{Q}_L X_\mu,
\end{equation}
which becomes, after rotating the quarks to the physical basis,
\begin{align}\label{eq:X-FCNC}
\L_{\rm X-FCNC} \simeq
& \frac{g_X}{3} \overline{\mathbf{u}}_L
\left(\begin{array}{ccc}
    V_{ub}^2           &  V_{ub}V_{cb}  & V_{ub} \\
    V_{ub}V_{cb} &  V_{cb}^2           & V_{cb}\\
    V_{ub}       &  V_{cb}      & 1
\end{array}\right)\gamma^\mu \mathbf{u}_LX_\mu
+ \frac{g_X}{3} \overline{\mathbf{d}}_L
\left(\begin{array}{ccc}
    0     & 0  & 0
    \\
    0     & 0  & 0    \\
    0 & 0  & 1
\end{array}\right)\gamma^\mu \mathbf{d}_LX_\mu.
\end{align}
The FCNC in the up sector induces flavor-changing top quark decays $t \to uX,cX$
which is presently not much constrained, and it contributes to $D^0-\bar {D^0}$ mixing and $D^+$ decays.
Note that the $D^0-\bar {D^0}$ mixing is doubly suppressed by the $V_{ub} V_{cb}$ factor and by the smallness of $g_X$.
We emphasize that there are no FCNC mediated by the $X$ gauge boson in the charged lepton sector, since the corresponding mass matrix
is diagonal.

\subsection{The scalar potential}

Now we turn our attention to the scalar sector of the model.  The most general renormalizable
scalar potential involving $\phi_1, \phi_2$ and $s$ that respects the symmetry of the model is given by
\begin{align}
  V &= m_{11}^2 (\phi_1^\dagger \phi_1) + m_{22}^2 (\phi_2^\dagger \phi_2) +
      m_{s}^2 s^*s + \frac{\lambda_1}{2} (\phi_1^\dagger \phi_1)^2 +
     \frac{\lambda_2}{2} (\phi_2^\dagger \phi_2)^2 +
     \lambda_3 (\phi_1^\dagger \phi_1)(\phi_2^\dagger \phi_2)\\
     &+
     \lambda_4 (\phi_1^\dagger \phi_2)(\phi_2^\dagger \phi_1) +
     \frac{\lambda_s}{2} (s^*s)^2 +
     \lambda_{1s} (\phi_1^\dagger \phi_1)(s^*s) +
     \lambda_{2s} (\phi_2^\dagger \phi_2)(s^*s) -
     \left[\mu (\phi_2^\dagger \phi_1)s + \rm{h.c.}\right].\nonumber
\end{align}
The presence of the $s$ field which allows for the cubic scalar coupling $\mu$ has several important consequences.
First, it removes an unwanted global symmetry and the associated pseudo-Goldstone boson
that would exist in its absence.  (The charge of  the $s$ field is chosen precisely to achieve
this.)  Second, the $\mu$ term allows to take the \emph{decoupling limit} of the model: by making $\mu\to\infty$, $v_s\to\infty$ and $m_{11}\to\infty$ (in order to keep $v_1$ finite), all extra scalars, the extra gauge boson, and the right-handed neutrinos can be made arbitrarily heavy, so
that the low energy theory is the SM.  Without this term, the masses of the second Higgs
doublet would have been bounded by about 600 GeV, analogous to the two Higgs doublet models with a
spontaneously broken discrete $Z_2$ symmetry~\cite{Branco:2011iw}. This decoupling behavior of $s$ enabled by $\mu$ is essential to evade
large deviations in $\Upsilon$ and $D^+$ decays, atomic parity violation and neutrino experiments.

The physical scalar spectrum consists of three neutral
scalars, one of which should be identified with the 125~GeV SM-like Higgs, a pseudoscalar, and a
charged scalar.  A pair of pseudoscalars and a charged scalar are absorbed by the $Z, X$ and $W^\pm$
gauge bosons.   The physical pseudoscalar boson mass is given by
\begin{equation}
  m_A^2=\mu\frac{v_1^2v_2^2+v_1^2v_s^2+v_2^2v_s^2}{\sqrt{2}v_1v_2v_s}.
\end{equation}
The charged scalar has a mass given
by
\begin{equation}
  m_{H^\pm}^2=\frac{1}{2}\lambda_4 v^2 + \mu\frac{ v_s v^2}{\sqrt{2}v_1 v_2},
\end{equation}
while the real scalar mass matrix is given by (in the basis $({\rm Re}(\phi_1), {\rm Re}(\phi_2), {\rm Re}(s)$)
\begin{equation}
  m_H^2=
\left(
\begin{array}{ccc}
 \lambda _1 v_1^2+\mu\frac{  v_2 v_s}{\sqrt{2} v_1} & (\lambda _3+\lambda _4)v_1 v_2-\frac{\mu  v_s}{\sqrt{2}} & \lambda_{1s}v_1 v_s-\mu\frac{  v_2}{\sqrt{2}} \\
  (\lambda _3+\lambda _4)v_1 v_2-\frac{\mu  v_s}{\sqrt{2}} & \lambda _2v_2^2 +\mu\frac{  v_1 v_s}{\sqrt{2} v_2} & \lambda_{2s}v_2 v_s -\mu\frac{  v_1}{\sqrt{2}}  \\
 \lambda _{1s}v_1 v_s-\mu\frac{  v_2}{\sqrt{2}} & \lambda _{2s} v_2 v_s-\mu\frac{  v_1}{\sqrt{2}} &  \lambda _sv_s^2+\frac{\mu  v_1 v_2}{\sqrt{2} v_s}
   \\
\end{array}
\right).
\end{equation}

Although it is not easy to write down simple analytic expressions for
the masses and mixings of the real scalars as  functions of the parameters
of the potential, we still can understand the interplay between the
mixing in the scalar sector and the symmetry structure of the model by
very simple arguments.  $\phi_2$ has diagonal couplings to
quarks and leptons which cannot distinguish between the ${\rm Re}(\phi_1)$ and ${\rm Re}(s)$ components of the
physical SM-like Higgs, $h$.  These couplings to fermions
have the structure $m_f/v_2\, \phi_2\bar{f} f$, and since $v_1^2+v_2^2=v^2$,
with $v\simeq 246~\GeV$, the Yukawa couplings are always larger compared to the
SM Yukawas.  For the top-quark Yukawa coupling to be in the perturbative range,
$v_2$ cannot be much smaller than $v$.  The scalar $\phi_1$ couples
off-diagonally to quarks (mediating flavor changing processes).  In order to have perturbative Yukawa
couplings with the top, $\tan\beta$ should lie in the range between
$0.5$ and $30$, with the upper limit arising from the off-diagonal Yukawa
coupling equal to $V_{cb}m_t/v_1$.

To understand the SM-like Higgs FCNC couplings, it is better to go to the Higgs basis, in which $H=c_\beta\phi_1+s_\beta\phi_2$ and $H'=-s_\beta\phi_1+c_\beta\phi_2$, which leads to $\vev{H}=v$, and $\vev{H'}=0$. Here, $H=(H^+,(h+v)/\sqrt{2})$. The mass matrix in the basis $({\rm Re}(H),{\rm Re}(H'),{\rm Re}(s))$, to leading order in each entry assuming $v\ll \mu,v_s\,$ is given by
\begin{equation}
\mathcal{M}^2\simeq\left(
\begin{array}{ccc}
 \frac{\left[\left(\lambda _2 t_{\beta }^2+2 \lambda _{34}\right) t_{\beta }^2+\lambda
   _1\right]v^2 }{\left(t_{\beta }^2+1\right){}^2}
   & \frac{t_{\beta } \left[\left(\lambda
   _{34}-\lambda _2\right) t_{\beta }^2+\lambda _1-\lambda _{34} \right]v^2}{\left(t_{\beta
   }^2+1\right){}^2} &
   \frac{\left[ \left(\lambda _{2s} t_{\beta }^2+\lambda
   _{1s}\right)v_s-\sqrt{2}  t_{\beta }\mu\right]v}{t_{\beta }^2+1} \\
   & \frac{\left(t_{\beta
   }^2+1\right) \mu  v_s}{\sqrt{2} t_{\beta }} &
   \frac{\left[2\left(\lambda _{1s}-\lambda _{2s}\right) t_{\beta }v_s
   	+\sqrt{2} (1-t_\beta^2)\mu \right]v }{2\left(t_{\beta }^2+1\right)} \\
   & &  \lambda _s v_s^2 \\
\end{array}
\right),
\end{equation}
where we have defined $\lambda_{34}=\lambda_3+\lambda_4$. The first entry is the SM-like Higgs state, the second is the flavor changing Higgs and the third refers to the state which does not couple to fermions. Integrating out the heavy scalars, when their masses are non-degenerate, yields the effective flavor changing operators
\begin{equation}\label{eq:higgs-eft}
  y_{ij}^{\prime u}\frac{H^\dagger H}{\Lambda^2}\bar Q_{iL} \tilde H u_{jR}
  	+y_{ij}^{\prime d}\frac{H^\dagger H}{\Lambda^2}\bar Q_{iL} H d_{jR},
\end{equation}
with
\begin{equation}
y^{\prime u,d}=y_t\left(
\begin{array}{ccc}
c_\beta m_u/m_t & 0 & -s_\beta V_{ub} \\
0 & c_\beta m_c/m_t & -s_\beta V_{cb} \\
0 & 0 & c_\beta
\end{array}
\right),
\end{equation}
a similar matrix for $y^{\prime d}_{ij}$, and also
\begin{align}
  \frac{1}{\Lambda^2}= & \frac{1}{(t_\beta^2+1)^2v_s^2}
  \Big( \frac{t_{\beta } \left(\lambda _{2,s}-\lambda _{1,s}\right)
   \left(\lambda _{1,s}+
   t_{\beta }^2 \lambda _{2,s}\right)}{\lambda _s^2}
    +\frac{\mu ^2
   \left(t_{\beta }-t_{\beta }^3\right)}{\lambda _s^2 v_s^2}\nonumber\\
   &
   + \frac{\mu
   \left(\left(t_{\beta }^2-3\right) t_{\beta }^2 \lambda _{2,s}+\left(3 t_{\beta
   }^2-1\right) \lambda _{1,s}\right)}{\sqrt{2} \lambda _s^2 v_s}
   +\frac{\sqrt{2} v_s t_{\beta }^2
   \left(\lambda _1-\lambda _{34}+\left(\lambda _{34}-\lambda _2\right) t_{\beta
   }^2\right)}{\mu  \left(t_{\beta }^2+1\right)}\Big).
\end{align}
This will induce top to charm Higgs decays, which will be analyzed in Sec.~\ref{sec:pheno}.

In this basis, the electroweak gauge bosons couple only to $H$,
 and hence any mixing of this state can only
reduce the couplings of the SM-like Higgs to $WW$ and $ZZ$.
The requirement that the SM-like Higgs boson couples
to the gauge bosons with strengths very close to the SM values
constrains the admixture of ${\rm Re}(H^0)$ with the other scalars.
LHC Higgs data constrain the sum of the square of these mixings to be about 0.1~\cite{Khachatryan:2016vau}.
LHC searches for a heavy Higgs boson decaying to
$ZZ$~\cite{CMS:2013ada,Aad:2015kna} are sensitive to masses roughly
between 200~GeV and 900~GeV, assuming production via gluon fusion and
a branching ratio to $ZZ$ similar to a SM-like Higgs of corresponding
mass. Due to the structure of the Yukawa couplings the heavy Higgs
bosons of the model have suppressed couplings to $tt$, leading to
smaller production cross sections, thus evading the LHC search limits.
(Note that in the large $\tan\beta$ limit, $h_{125 {\rm GeV}}\sim {\rm Re}(H^0)\sim {\rm Re}(\phi_2^0)$, and since
only $\phi_2$ has a $tt$ coupling, the couplings of all heavy Higgs bosons with
$tt$ will be suppressed by small mixing angles.)
Besides, due to the $X-Z$ mixing, the real component of $H_1$ will couple to $X$ like
\begin{equation}\label{eq:h-goldstone}
  \mathcal{L}_{hXX}=\frac{g_X^2}{9}\frac{v_1^2v_2^2}{v^3}{\rm Re}(H^0)X_\mu X^\mu.
\end{equation}
This coupling will contribute mainly to the invisible width of the Higgs, as we will see in the next section.

With the aid of the cubic scalar coupling $\mu$ the mass of the charged
scalar can be raised above the electroweak scale, which may be very important for the following reason.  In
type-II 2HDM, where each Higgs couples exclusively to up- and
down-type quarks, the charged Higgs contribution to $b\to s\gamma$
transitions constrains its mass to be above $\sim 400-500~\GeV$ for
$\tan\beta\simeq 1$~\cite{Hermann:2012fc}. Although our model is
not a type-II 2HDM, the $\bar t\, b H^+$ and $\bar t s H^+$
 couplings are similar, and therefore a comparable bound should be applicable here
 as well.
 \footnote{As a
  side remark, we note that the $\mu$ parameter cannot be made
  arbitrarily large while keeping the Higgs mass light, as it would
  violate unitarity in certain scattering processes.  The amplitude for the scattering
  $\phi_i \phi_h \rightarrow \phi_i \phi_j$ would grow
  like $\mu^2/m^2$, where $m$ is the mass of the virtual scalar
  exchanged, which would violate unitarity if $\mu \gg m$.}
  LHC searches for $H^\pm\to t b$~\cite{Aad:2015typ} are
sensitive to masses below $250-300$~GeV only if $\tan\beta>2$.
 As an example, the parameters $\tan\beta=10$, $v_s=300~\GeV$,
$\mu=181~\GeV$, $\lambda_1=1$, $\lambda_2=0.24$, $\lambda_s=2$,
$\lambda_3=0.1$, $\lambda_4 = 1.5$, $\lambda_{1s}=1$, and
$\lambda_{2s}=0.1$ lead to a physical Higgs at $125~\GeV$ with
couplings almost identical to the SM Higgs (except for small flavor
violating couplings to $ut$ and $ct$), while the two scalars, the
pseudoscalar and the charged one would have masses of 620~GeV,
420~GeV, 620~GeV, and 590~GeV. This scalar spectrum would lead to a
small deviation on the electroweak $T$ parameter of about $\Delta
T=0.13$.

\section{Phenomenology: key constraints}
\label{sec:pheno}

The phenomenology of a light mediator coupled to the standard model
fields through kinetic mixing has been studied in the literature in
great detail (see Ref.~\cite{Essig:2013lka} and references
therein). Our model has a very rich phenomenology as, besides mixing kinetically
with the photon, the $X$ gauge boson also mixes with the $Z$ via mass terms.
Furthermore, the couplings of $X$ to fermions are flavor non-universal, which would lead
to flavor changing neutral currents mediated by both $X$ and the new scalar
bosons needed for symmetry breaking.   In this section we present the main
results obtained from various constraints arising from low energy processes.
For definiteness, when quoting numbers we focus on benchmark points
where we set $\varepsilon=0$ and $\tan\beta=0.5,2$, while in presenting the
constraints as plots we scan the entire allowed range of $\tan\beta = (0.5,25)$,
with $\varepsilon=0$. We present in Table \ref{tab:bounds1} a summary of
the most constraining experimental limits together with a brief description of each
bound. The branching ratios of $X$ are shown in Fig.~\ref{fig:BRs}, while in  Figs.~\ref{fig:bound-big}, \ref{fig:bounds} and \ref{fig:tanb} we present
a summary of the most relevant constraints.  Additional experimental constraints are analyzed in Sec.~\ref{sec:appendix},
which turn out to be important, but only to a lesser degree.  We elaborate now on how the main results
summarized in Table \ref{tab:bounds1} and Figs.~\ref{fig:bound-big}, \ref{fig:bounds} and \ref{fig:tanb} are obtained.

\subsection{Branching ratios of $X$}

\begin{figure}[!t]
 \begin{center}
   \includegraphics[scale=1]{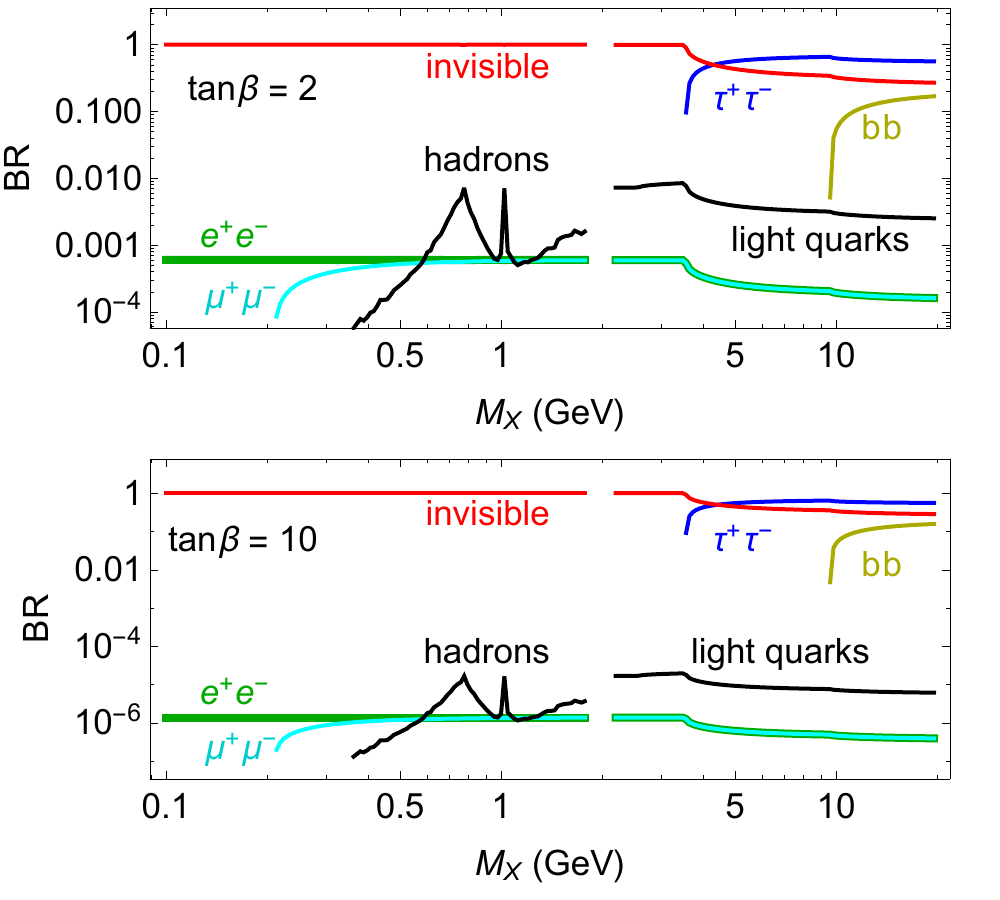}
 \end{center}
 \caption{Branching ratios of $X$ for two values of $\tan\beta\equiv v_2/v_1$ with no kinetic mixing. }
 \label{fig:BRs}
\end{figure}

Before discussing the constraints in detail, we first explore the $X$ branching ratios which will
define the typical signature of the new gauge boson. If $M_X$ is lighter than the tau mass, it can
only decay to first and second family charged fermions, and to all neutrinos.
In this case, the partial widths to the charged fermions go as $\sim g_X^2/(1+t_\beta^2)^2$ while the
width to $\nu_\tau\nu_\tau$ goes as $g_X^2$, and hence the branching ratio to the first two families
has a $t_\beta^{-4}$ suppression (in the limit of large $t_\beta$). For instance, if $M_X<2m_\tau$, we obtain
\begin{equation}
  {\rm BR}(X\to e^+e^-)=\frac{1-4s_w^2+8s_w^4}{7-4s_w^2+8s_w^4+12t_\beta^2+9t_\beta^4}
  	= \frac{0.056}{0.72+1.3t_\beta^2+t_\beta^4}.
\end{equation}
In Fig.~\ref{fig:BRs} we provide the
exact branching ratios of $X$ for two different values of $t_\beta$.

To obtain the hadronic partial width for $M_X$ below 1.8~GeV we use the experimentally
measured ratio
\begin{equation}
  R(s)=\frac{\sigma(e^+e^-\to {\rm hadrons};s)}{\sigma(e^+e^-\to \mu^+\mu^-;s)},
\end{equation}
where $s$ is the center of mass energy of the $e^+e^-$ collision~\cite{Whally:2003, hepdata}. We
estimate the $X$ hadronic width to be~\footnote{In fact, the $X$ branching ratios should not be
exactly the values obtained here. The hadronic cross section at low energy $e^+e^-$ colliders is
dominated by photon exchange. Since the coupling of $X$ to light quarks arrives from $X-Z$
mixing, they differs from the photon couplings: they are not universal and have an axial-vector
component. Nevertheless, the hadronic branching ratios derived here are expected to provide a
good approximation to the exact ones (which cannot be calculated perturbatively).}
\begin{equation}
  \Gamma(X\to {\rm hadrons})=\Gamma(X\to \mu^+\mu^-) R(s=M_X^2).
\end{equation}
Above $2.2~\GeV$ we calculate the partial widths to partons.

\subsection{Lepton universality in $\Upsilon$ decays}

Precise measurements of the $\Upsilon\to\tau^+\tau^-$ and
$\Upsilon\to\mu^+\mu^-$ branching ratios by
BaBar~\cite{delAmoSanchez:2010bt} constrain the deviation from lepton
universality via the ratio
\begin{equation}
  R_{\tau\mu}\equiv\frac{\Gamma(\Upsilon(1S)\to\tau^+\tau^-)}{\Gamma(\Upsilon(1S)\to\mu^+\mu^-)}
  = 1.005\pm0.013(stat.)\pm0.022(syst.)\,.
\end{equation}
As the $X$ boson couples dominantly to the third family, this
measurement can be used to constrain $g_X$. In the limit of small $Z-X$ mixing   and neglecting the tiny $Z$ exchange diagram, we
obtain
\begin{equation}
  R_{\tau\mu}\simeq 1-2\frac{g_X^2}{e^2}\frac{M_\Upsilon^2}{M_\Upsilon^2-M_X^2},
\end{equation}
where the second term comes from the $\gamma-X$ interference. In our numerical evaluation we used the exact expression for $R_{\tau\mu}$. This imposes $g_X<0.027$ for $m_X\ll m_\Upsilon$. If $m_X\gg
m_\Upsilon$, this process actually constrains $v_s$. In such case,
$v_s>960~\GeV$, roughly independent of $\tan\beta$.

\subsection{$\Upsilon \rightarrow X \gamma$ decay}

The decay $\Upsilon\to X_L \gamma$ can also occur and can be used
to constrain the parameters of the model.\footnote{We have
  checked that $\Upsilon\to X_LX_L$ does not lead to any meaningful
  bound due to a weaker experimental limit on the branching fraction.}
  Here $X_L$ is the longitudinal mode of $X$. Although this process
involves gauge bosons, the equivalence theorem tells us that this
width is actually probing the Yukawa coupling of the corresponding
Goldstone to the $b$ quarks, and therefore the bound is independent of
whether the theory is gauged or not,  as long as $M_X\ll m_b$ holds.  Yang's theorem,
which states that a vector particle cannot decay into a pair of
massless spin-1 particles, does not apply in this case as the
$\Upsilon$ is decaying into the longitudinal mode of $X$ and a
massless photon. Moreover, due to charge conjugation symmetry, only the
axial-vector coupling of $X$, that is, $c_{bR}-c_{bL}$ from
Eq.~(\ref{eq:Xcoupling}), will contribute to $\Upsilon\to X_L \gamma$.
This branching ratio can be computed using non-relativistic
effective field theory \cite{Manohar:2003xv}, where the
amplitude is approximated by the zero momentum amplitude for the hard
scattering times the wave function of the $\Upsilon$ at the origin,
$A_\Upsilon\simeq A(0)\psi(0)$. We get rid of the wave function at the
origin by taking the ratio of this width with a measured decay width
like $\Upsilon\to e^+ e^-$. Therefore we have
\begin{align}
  R \equiv \frac{{\rm BR}(\Upsilon\to X_L \gamma)}{{\rm BR}(\Upsilon\to e^+e^-)}
     &= \frac{|\psi(0)|^2\left|A(0;b\bar b\to X_L\gamma)\right|^2}{|\psi(0)|^2\left|A(0;b\bar b\to e^+e^-)\right|}\simeq \frac{2g_X^2v_1^4m_b^2}{9e^2v^4 M_X^2} \nonumber\\
     &= \frac{2m_b^2v_1^4}{e^2v^2(v^2v_s^2+v_1^2v_2^2)}<\frac{4.5\times 10^{-6}}{0.0238},
\end{align}
where the right-hand side of the inequality shows the measured values
of the branching ratios being considered~\cite{Agashe:2014kda}.  The
constraint on $v_s$ is $v_s>2(0.5)~\TeV$ for $\tan\beta=0.5(2)$.

\subsection{$D^0-\overline{D^0}$ mixing}

A light gauge boson with flavor changing couplings to quarks can contribute to meson-antimeson mixing.
In our model, since the first two families carry no $U(1)_{B-L}^{(3)}$ charge, and since
the third family quark mixings arise from the up-quark mass matrix, these constraints are not severe.
The effective interaction mediated by the $X$ gauge boson responsible for $D^0-\overline{D^0}$ mixing can
be written as (see Eq. (\ref{eq:X-FCNC}))
\begin{equation}
{\cal L}_{\rm eff} = C(q^{2}) (\overline{u}_L\gamma_\mu c_L)^2,
\end{equation}
where
\begin{equation}
C(q^{2}) = \frac{g_X^2}{9} \frac{|V_{ub}V_{cb}|^2}{q^2-M_X^2}~.
\end{equation}
Here $q^2$ represents the momentum transfer. Demanding that the
new contribution does not exceed the experimental value of $\Delta m_D$, a limit on $C(m_D^2)$ has been
obtained to be \cite{Blum:2009sk}
\begin{equation}
C(m_D^2) < \frac{5.9 \times 10^{-7}}{{\rm TeV}^2}~.
\end{equation}
For the case of a light $X$, this constraint leads to a limit $g_X < 2.6 \times 10^{-2}$, which is significant,
but within our range for $g_X$.  When the $X$ boson mass is much larger than $m_D$, the limit becomes
$g_X < 1.4 \times 10^{-2}M_X/{\rm GeV}$.  The limit is plotted in Fig. \ref{fig:bounds} for the full range of $M_X$.

The Higgs bosons in the model also mediate $D^0-\overline{D^0}$ mixing.
 The contribution from
tree level neutral scalar exchange to meson oscillations, in general,
can be written as~\cite{Babu:2009nn, Babu:1999me,Golowich:2009ii}
\begin{align}
  (\Delta m_S)_\varphi=\frac{1}{3}\frac{f_S^2 m_S B_S}{m_\varphi}&\left\{
      \left[\frac{1}{6}\frac{m_S^2}{(m_{qi}+m_{qj})^2} +\frac{1}{6}\right]
           {\rm Re}\left(h_{ij}+h_{ji}^{{*}}\right)^2 \right.\nonumber\\
&\left.-\left[\frac{11}{6}\frac{m_S^2}{(m_{qi}+m_{qj})^2}+\frac{1}{6}\right]
           {\rm Re}\left(h_{ij}-h_{ji}^{{*}}\right)^2\right\},
\label{eq:meson}
\end{align}
where $f_S$ is the meson decay constant, $B_S$ is the bag parameter,
$m_S$ is the meson mass, $h_{ij}$ and $m_\varphi$ are the couplings to
and masses of the physical scalars, and $m_{qi,j}$ are the masses of
the quarks constituting the meson. Since the flavor structure is determined, we obtain
\begin{equation}
  \Delta m_{D}^{\rm scalars}=-2.4\times10^{-10}\left(\frac{100\,\GeV}{m_\varphi}\right)^2
         {\rm Re}\left(\frac{h_{12}^u}{\sqrt{2}m_c/v}\right)^2{\rm GeV},
\end{equation}
which should be smaller than the theoretical uncertainty of $2.7\times
10^{-15}~\GeV$~\cite{Agashe:2014kda}. As $h_{12}^u\sim
\sqrt{2}V_{ub}V_{cb}m_c/v \sim 2 \times 10^{-6}$, the new scalar
contributions are within experimental limits, even with the heavy
Higgs boson mass $m_\varphi$ being of order 100 GeV.

\begin{table}[t!]
  \begin{center}
    \begin{tabular}{| C{4cm} | L{11.8cm} |} \hline

Experimental constraint & Remarks \\ \hline\hline
$K^+$ decays & Enhanced $K^+\to\pi^+\nu\bar\nu$ branching ratio at loop level \\ \hline
$B^+$ decays & Enhanced $B^+\to K^+\nu\bar\nu$ branching ratio at loop level. It shows a strong dependence on the mass of the charged scalar \\ \hline
Neutrino oscillations & Non-universal matter effects bounded by atmospheric neutrinos\\ \hline
 Atomic parity violation & $X-Z$ mixing modifies weak charge of $^{133}$Cs \\ \hline
 $\Upsilon$ decay & $\Upsilon \rightarrow \gamma X \rightarrow \gamma \nu \bar{\nu}$: Goldstone boson
 equivalence theorem constrains Yukawa coupling \\ \hline
 $\Upsilon$ decay & $\Upsilon \rightarrow \tau^+\tau^-$: Direct constraint on the gauge coupling as the process only involves third family fermions \\ \hline
 Electroweak $T$ parameter & $Z-X$ mixing modifies $M_Z/M_W$ and constrains the mixing parameter $s_X$ \\ \hline
 $D^0-\overline{D^0}$ mixing & Mediated by scalar constrains mass of heavy scalar $> O(100)$ GeV; significant
 constraint on the coupling of $X$ only when $X$ mass is below or close to the $D^0$ mass\\ \hline
 $D^+$ decays & $D^+\to\pi^+ X$ contributes to the total $D^+$ width and to the $\pi^+\ell^+\ell^-$ branching ratio. When the equivalence theorem is valid, this process probes the Yukawa coupling\\ \hline
    \end{tabular}
  \end{center}	
  \caption{A summary of the major experimental constraints on the model.}
\label{tab:bounds1}
\end{table}

\subsection{$D^+\to \pi^+ e^+ e^-$ and $D^+$ lifetime}

The flavor properties of $X$ can contribute to the $D^+\to\pi^+ X \to \pi^+ e^+ e^-$ branching ratio which is bounded to be below $1.1\times 10^{-6}$~\cite{Agashe:2014kda}. This process can be better understood by use of the equivalence theorem, where the Goldstone coupling to $uc$ is given in Eq.~(\ref{eq:goldstone-fermions}). The $D^+$ to $\pi^+$ transition can be parametrized by the form factors
\begin{equation}
  \langle \pi^+(p_2)|\bar{u}\gamma_\mu c|D^+(p_1)\rangle=F_+(q^2)(p_1+p_2)_\mu+F_-(q^2)(p_1-p_2)_\mu.
\end{equation}
At low recoils (for $M_X\ll M_{D^+}$), the transition comes entirely from $F_+$, which can be determined by use of chiral perturbation theory for heavy hadrons (see e.g. Ref.~\cite{Burdman:2003rs}),
\begin{equation}
  F_+(s)=\frac{f_D}{f_\pi}\frac{g_{D^*D\pi}}{1-s/M_{D^*}^2}.
\end{equation}
Here, $f_D=200~\MeV$ and $f_\pi=130~\MeV$ are the $D^+$ and $\pi^+$ decay constants, and $g_{D^*D\pi}=0.59$ is the strong coupling of $D^*\to D\pi$ decay, all yielding $F_+(0)=0.91$. Numerically, this form factor agrees with the one obtained by assuming vector meson dominance~\cite{Babu:1987xe}.
The $D^+\to\pi^+ X$ partial width is then given by
\begin{equation}
  \Gamma(D^+\to\pi^+ X)=\frac{1}{144\pi}|F_+(M_X^2)|^2g_X^2|V_{ub}|^2|V_{cb}|^2\frac{m_{D^+}^3}{M_X^2},
\end{equation}

Not requiring the $e^+e^-$ pair in the final state makes very hard to reconstruct the $D^+$ meson as $X$ will typically decay to neutrinos (see Fig.~\ref{fig:BRs}). Nevertheless, one can still constrain the model with the total $D^+$ width. As a conservative requirement, we demand that the partial width $D^+\to\pi^+ X$ does not exceed the $D^+$ total width minus the partial inclusive width to $K^0$ and $\bar K^0$ (to which this new decay does not contribute), that is $\Gamma(D^+\to\pi X)<0.39\,\Gamma_{D^+}$~\cite{Agashe:2014kda}.
This constraint is included in our numerical analysis.

\begin{figure}[!t]
 \begin{center}\vspace{-1.3cm}
   \hspace{-1.cm}\includegraphics[scale=1.05]{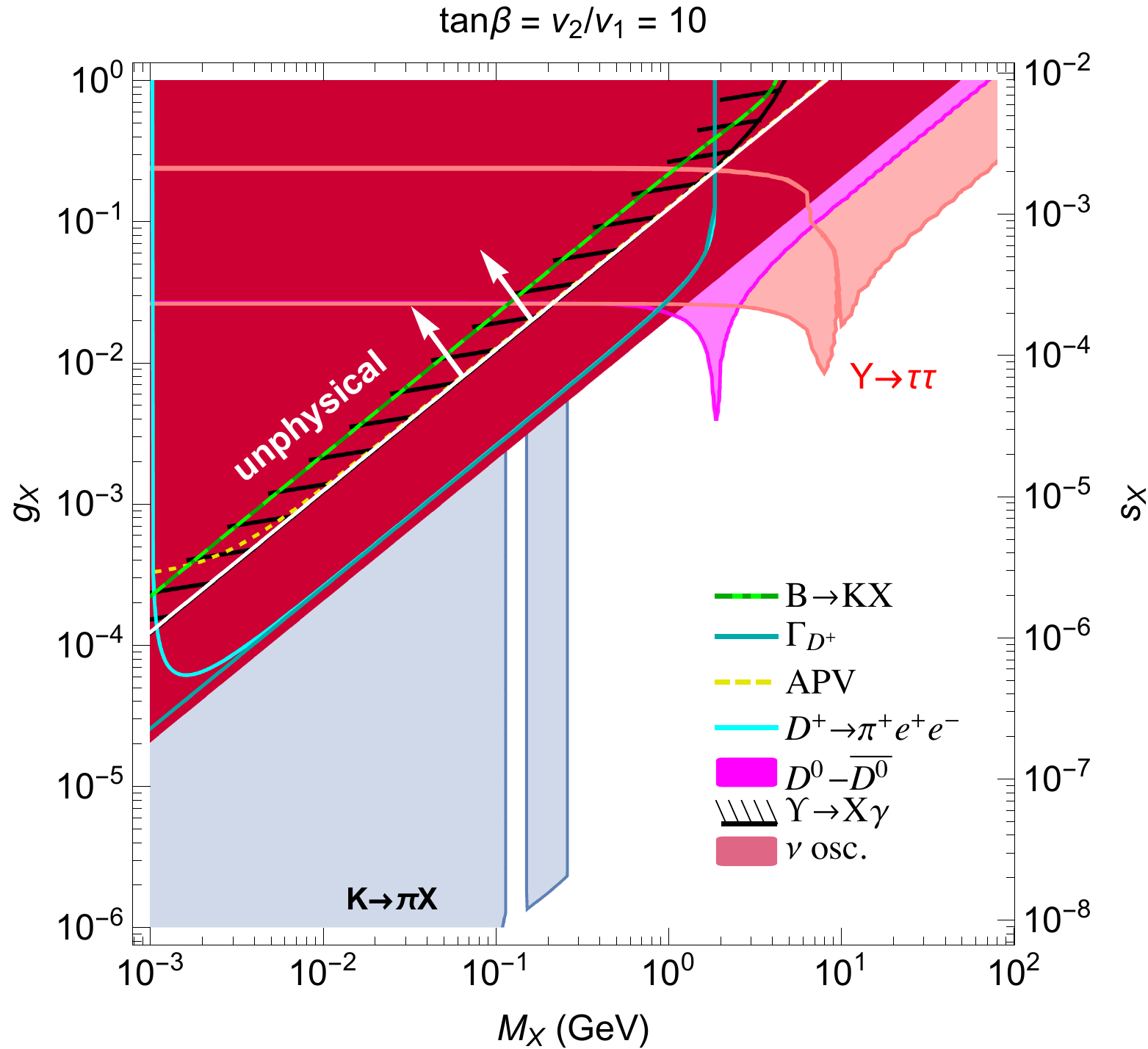}
 \end{center} \vspace{-0.7cm}
 \caption{Constraints on the $U(1)_{B-L}^{(3)}$ gauge boson mass
   $M_X$ and coupling $g_X$ for $\tan\beta=10$. For
   convenience the $X-Z$ mixing, $s_X$, is also shown.  Notice that
   for a given $g_X$, the mass of the gauge boson $M_X$ is bounded
   from below, so there is an unphysical region in the upper left
   corner of the $M_X\times g_X$ plane (delineated by the white
   line). The ``$\nu$ osc.'' bound
   comes from non standard interaction effects (matter potential) on
   atmospheric neutrinos. ``APV'' refers to atomic parity violation. 
   Here the charged Higgs mass, relevant to the $B\to K X$ constraint, is taken to be 1200~GeV.}
 \label{fig:bound-big}
\end{figure}

\subsection{$K^+\to\pi^+X$ and $B^+\to\pi^+ X$}
Although the flavor changing couplings in the down-quark sector can be put to zero, one loop
corrections will still generate a non-negligible amount of flavor changing. Kaon and $B$ decays are
particularly sensitive if the $X$ boson is below the meson mass. More specifically, the loop
corrections can contribute to the $K^+ \to\pi^+X\to\pi^+\nu\bar\nu$ ($B^+ \to K^+X\to K^+\nu\bar\nu$) branching ratio which is measured to be about $10^{-10}$~\cite{Anisimovsky:2004hr, Artamonov:2008qb} ($1.6\times 10^{-5}$~\cite{Agashe:2014kda}). We will discuss  the Kaon decay in detail and the results can be promptly generalized for the $B$ decay. As the longitudinal mode of $X$ dominates the contribution, we are interested in the one loop coupling
$g_{sdX}(\partial_\mu G_X)\bar{s}\gamma^\mu d $. This calculation differs from the usual $Z$-induced Kaon decay precisely by the dominance of the longitudinal mode, which lead us to the following considerations.
Since the internal $X$ vertex effectively couples to the Yukawa instead of the gauge coupling, 
we can safely take all quark masses, except for the top, to be zero.
The charm quark contribution to the amplitude is suppressed in our scenario, and thus we neglect it.
Moreover, the usual counterterms from the self-energy diagrams are omitted since they are proportional to the mass of the $s$ or the $b$.

\begin{figure}[!t]
 \begin{center}
   \includegraphics[scale=0.2]{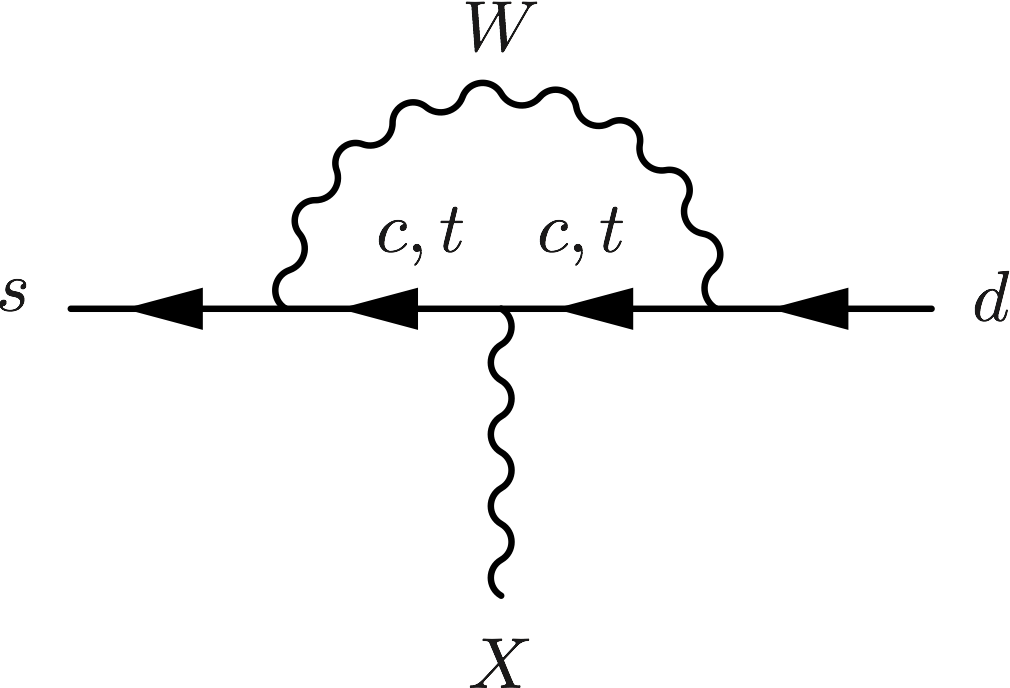}\hspace{0.1cm}
   \includegraphics[scale=0.2]{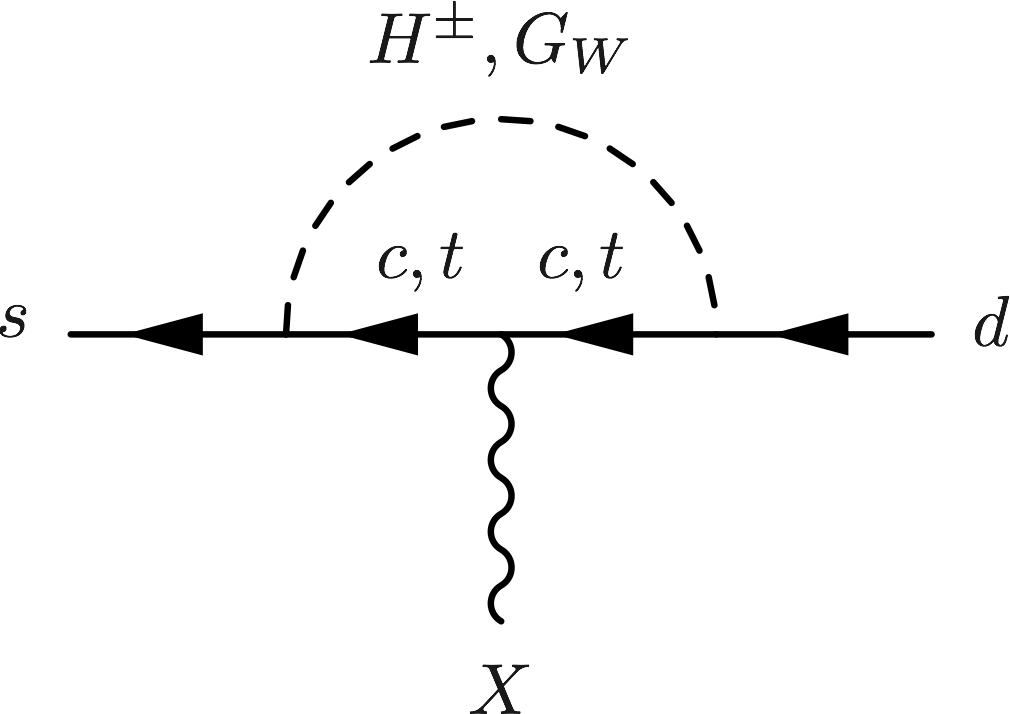}\hspace{0.1cm}
   \includegraphics[scale=0.2]{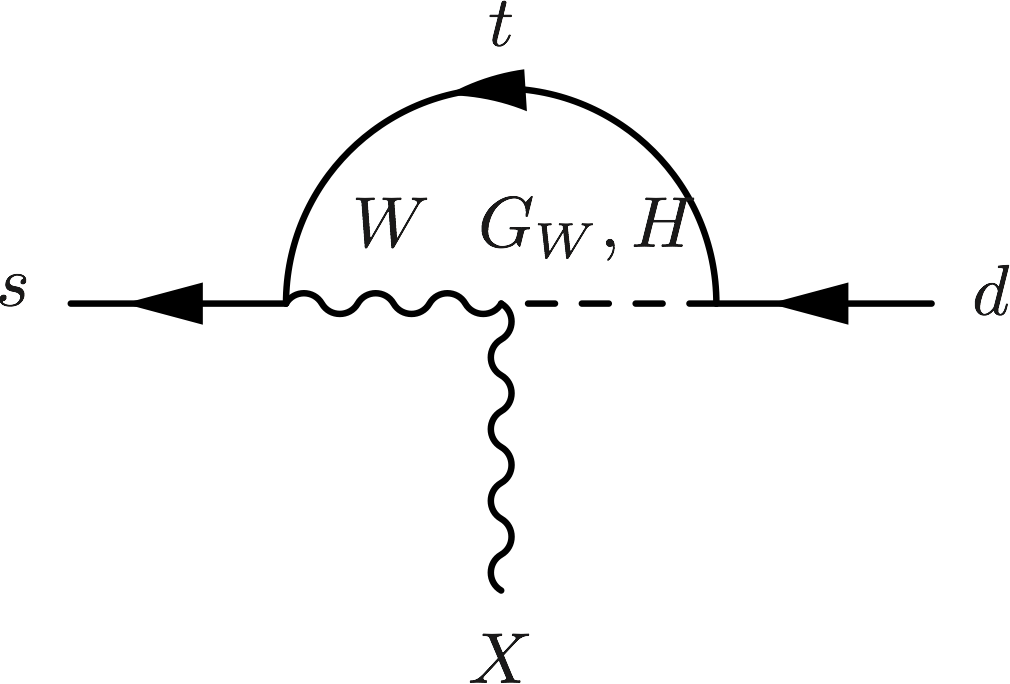}\hspace{0.1cm}
   \includegraphics[scale=0.2]{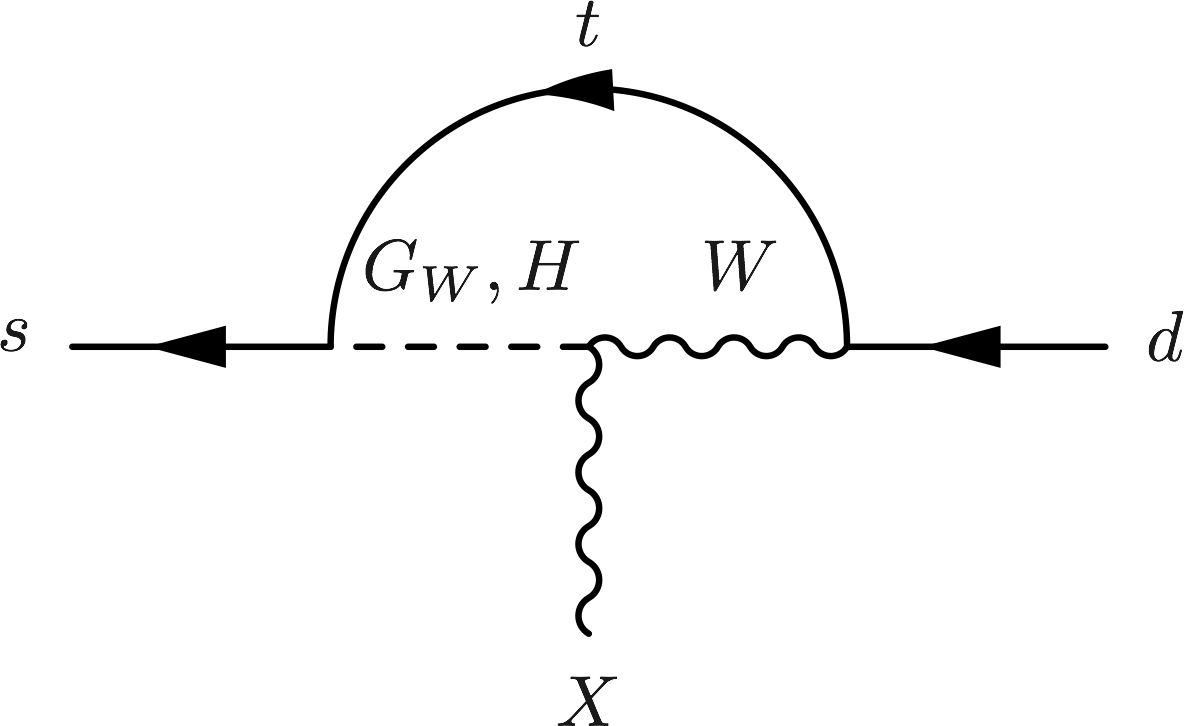}\\
   {\footnotesize \hspace{-0.5cm} (a) \hspace{3.3cm} (b) \hspace{3.cm} (c)\hspace{3.5cm} (d)}
 \end{center} \vspace{-0.3cm}
 \caption{Feynman diagrams involved in the calculation of $K^+\to\pi^+ X$. Analogous diagrams were computed for $B^+ \to K^+ X$.}
 \label{fig:diagrams}
\end{figure}

There are three main contributions (in the Feynman gauge) to this coupling, i.e., loops with transverse $W$, longitudinal $W$ or charged Higgs, and both transverse $W$ and charged Higgs (via a $W^\pm H^\mp G_X$ coupling), see Fig.~\ref{fig:diagrams}. For the longitudinal $W$ diagram in Fig.~\ref{fig:diagrams}(a), the internal fermions could be $tt$, or a top and a light up-type quark. These contributions scale as (see Eq.~(\ref{eq:goldstone-fermions}))
\begin{equation}
 g_{sdX}^{(1)}\sim\frac{g^2g_X}{96\pi^2M_X} \times \{\,V_{td}V_{ts}^*c_\beta^2\,\,,\,\,V_{td}(V_{cb}V_{cs}^*+V_{ub}V_{us}^*)\,\}\sim (1.5-0.6i) 10^{-7}\frac{g_X}{M_X}\times \{-c^2_\beta\,\, ,\,\,1\}.
\end{equation}
For the longitudinal $W$ and the charged Higgs in Fig.~\ref{fig:diagrams}(b), having a light quark in the loop would suppress the diagram by $m_{\rm light}^2/m_{t}^2$, so these contributions are negligible. Thus, the top loop exchange goes as
\begin{equation}
g_{sdX}^{(2)}\sim\frac{g^2g_X}{96\pi^2M_X} V_{td}V_{ts}^*\frac{c_\beta^2}{s_\beta}\sim -(1.5-0.6i) 10^{-7}\frac{g_X}{M_X} \frac{c_\beta^2}{s_\beta}.
\end{equation}
Finally,  for the diagram arriving from the $W^\pm H^\mp G_X$ coupling in Fig.~\ref{fig:diagrams}(c) and \ref{fig:diagrams}(d), only an internal top will lead to sizable contributions,
\begin{equation}
g_{sdX}^{(3)}\sim\frac{g^2g_X}{96\pi^2M_X} V_{td}V_{ts}^*c_\beta\sim -(1.5-0.6i) 10^{-7}\frac{g_X}{M_X}c_\beta.
\end{equation}
We emphasize that all contributions are comparable and have slightly different dependences on $t_\beta$, which will result in a bound from $K^+\to\pi^+ X$ that depends mildly on $t_\beta$.

\begin{figure}[!h]
 \begin{center}
   \includegraphics[scale=0.63]{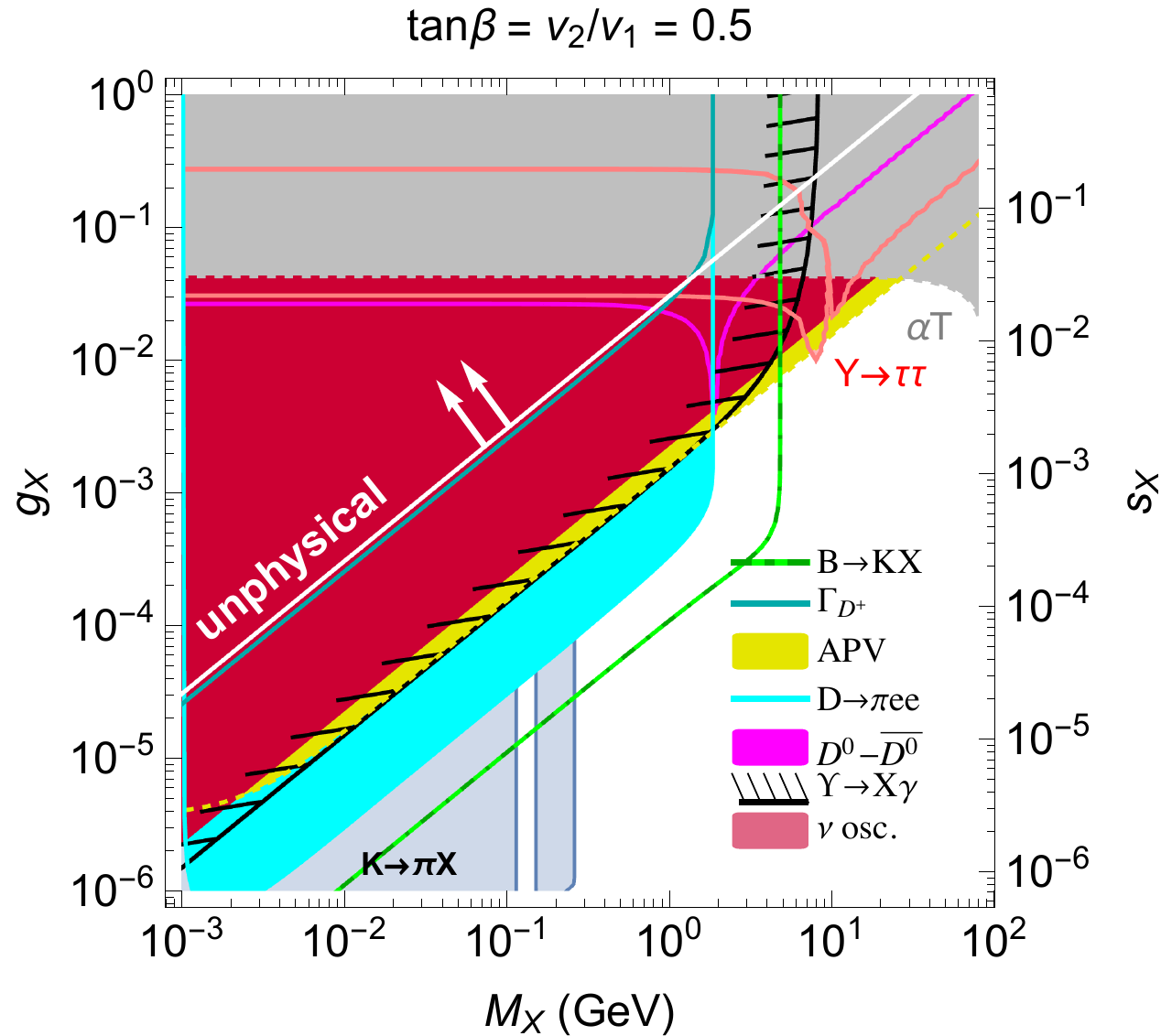}
   \includegraphics[scale=0.63]{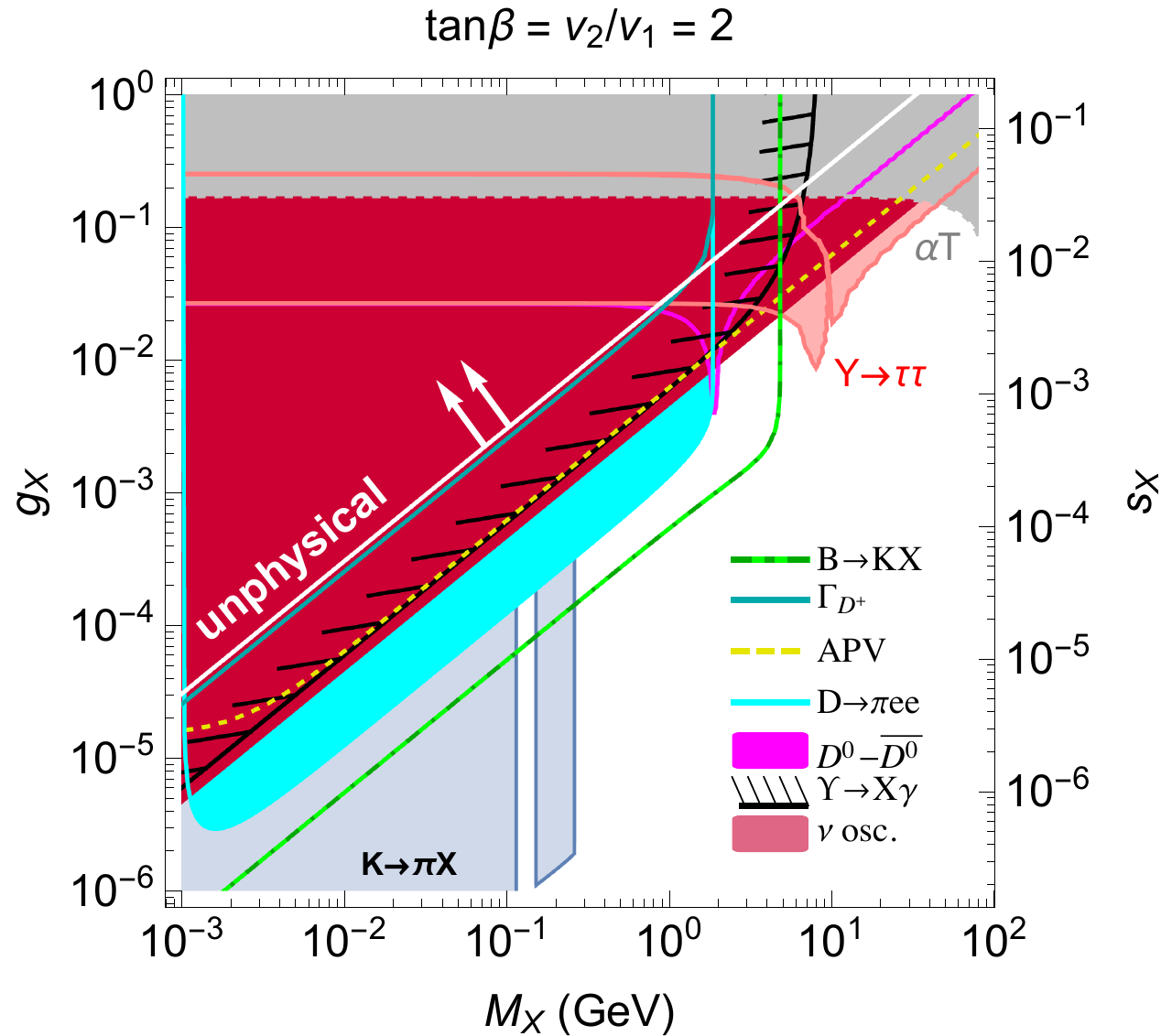}\\
   \includegraphics[scale=0.63]{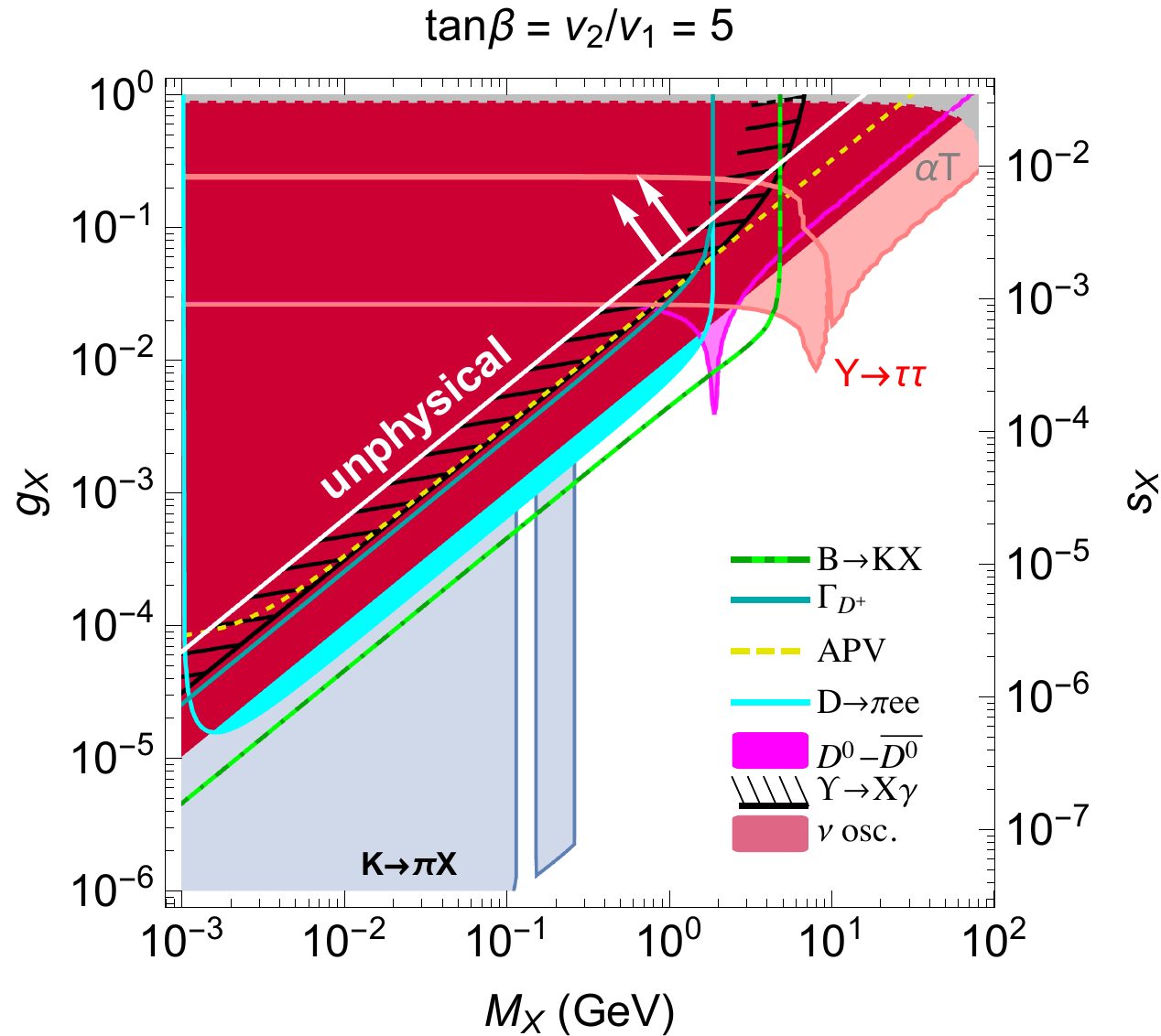}
   \includegraphics[scale=0.63]{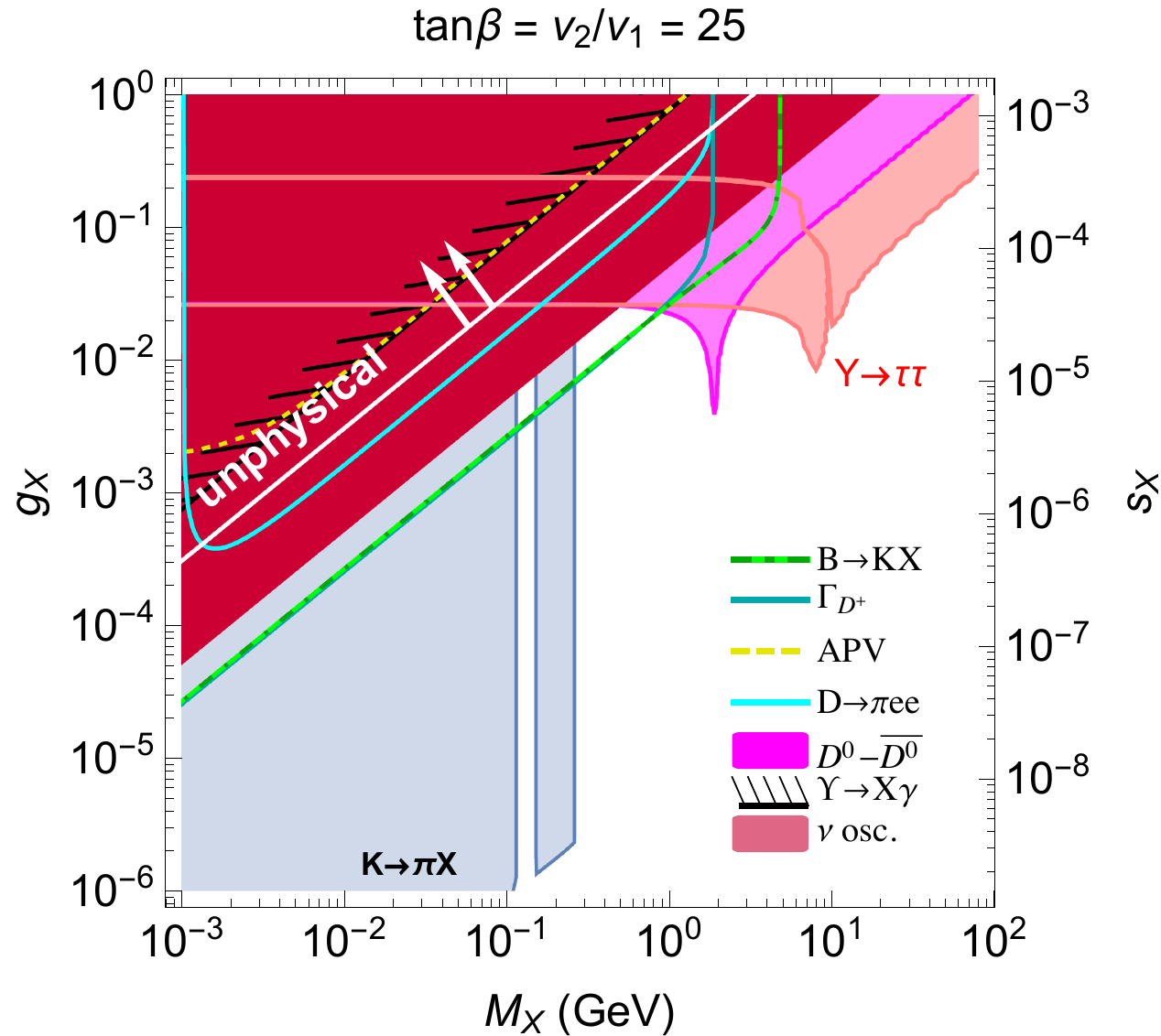}
 \end{center} \vspace{-0.7cm}
 \caption{Constraints on the $U(1)_{B-L}^{(3)}$ gauge boson mass
   $M_X$ and coupling $g_X$ for $\tan\beta=0.5,2,5,25$. For
   convenience the $X-Z$ mixing, $s_X$, is also shown.  Notice that
   for a given $g_X$, the mass of the gauge boson $M_X$ is bounded
   from below, so there is an unphysical region in the upper left
   corner of the $M_X\times g_X$ plane (delineated by the white
   line). The ``$\nu$ osc.'' bound
   comes from non standard interaction effects (matter potential) on
   atmospheric neutrinos. ``APV'' refers to atomic parity violation.}
 \label{fig:bounds}
\end{figure}

A full calculation of these loop amplitudes yields the following result (for similar calculations see e.g. Refs.~\cite{Inami:1980fz, Hall:1981bc, Frere:1981cc, Freytsis:2009ct, Davoudiasl:2012ag})
\begin{align}
  &g_{sdX}=i\frac{g^2g_X}{96\pi^2M_X}(T_1+T_2+T_3),\\ \label{eq:t1-kaon}
  &T_1=\frac{2t}{(t-1)^2}V_{td}\left[-(V_{cb}V_{cs}^*+V_{ub}V_{us}^*)(t-1)\log t+c_\beta^2V_{ts}^*(t-1-\log t)\right],\\
  &T_2=-\frac{V_{td}V_{ts}^*\,t\,c_\beta^2}{(t-1)^2(u-1)^2s_\beta}\left[s_\beta t(u-1)^2(1-t+\log t)+c_\beta u(t-1)^2 (1-u+\log u)\right],\\ \label{eq:t3-kaon}
  &T_3=\frac{4  V_{td}V_{ts}^* \,t\,u \,c_\beta}{(t-1)(u-1)(t-u)}\left[(t-1)\log u-(u-1)\log t\right],
\end{align}
with $t=m_t^2/M_W^2$ and $u=m_t^2/M_{H^\pm}^2$. The $T_{1,2,3}$ terms correspond to the
loop diagrams containing a transverse $W$, $G_W^\pm$ and $H^\pm$, and the triple coupling
$W^\pm H^\mp G_X$, respectively.
The amplitude is given by
\begin{equation}
  A(K^+\to\pi^+ X_L)=g_{sdX}\langle \pi | \bar d \gamma_\mu s | K\rangle q^\mu \simeq 2 g_{sdX}F_+(0)p\cdot q,
\end{equation}
where the form factor $F_+(0)=0.96$~\cite{Marciano:1996wy}. This leads to the partial width
\begin{equation}
  \Gamma(K^+\to\pi^+X_L)\simeq\frac{1}{16\pi^2}|g_{sdX}|^2 |F_+(0)|^2 M_K^3\left(1-\frac{M_\pi^2}{M_K^2}\right)^3,
\end{equation}
where we have neglected $M_X^2/M_K^2$ terms.

The two best experimental measurements of $K^+\to\pi^+\nu\bar\nu$ have different cuts for the pion
momentum. In Ref.~\cite{Anisimovsky:2004hr}, the pion momentum is required to be between 211
and 229~MeV, and the measurement yielded ${\rm BR}(K^+\to\pi^+\nu\bar\nu)=(1.47^{+1.30}_{-0.89})
\times 10^{-10}$, while in Ref.~\cite{Artamonov:2008qb} the pion momentum is required to be between
140 and 199~MeV and the measurement reads ${\rm BR}(K^+\to\pi^+\nu\bar\nu)=(1.73^{+1.15}
_{-1.05})\times 10^{-10}$. These cuts in momentum translate into the two intervals
$M_X<114~\MeV$ and $151<M_X<260~\MeV$, where the constraint should be valid.
The standard model value for this branching ratio is $(0.80\pm0.11)\times 10^{-10}$. The constraint
is shown in Figs.~\ref{fig:bound-big} and \ref{fig:bounds}, where we required the sum of the standard and new contributions not to exceed the $2\sigma$ experimental value. The gap in the excluded region is the result of the two intervals for $M_X$. 

For the $B^+\to K^+\nu\nu$ decay, a very similar calculation is performed and yields a bound that is weaker than the Kaon decay bound, but goes to higher values of $X$ masses~\footnote{For the $B\to K$ transitions, the relevant form factor is smaller, $F_+(0)=0.331$~\cite{Ball:2004ye}.}. Furthermore, the dependence with the mass of $H^+$ and $\beta$ is more pronounced in the $B$ decay constraint. The reason is because all contributions in Eq.~(\ref{eq:t1-kaon}) are comparable for $K^+\to\pi^+X$, but only the last one is significant for $B^+\to K^+X$,  and thus the $\beta$ dependence and the interplay with Eq.~(\ref{eq:t3-kaon}) can lead to cancelations. We have checked numerically that e.g. for $\tan\beta=5(10)$ such cancelation is possible by having the charged Higgs mass in the range
$600<M_{H^+}<850~\GeV$ ($1<M_{H^+}<1.5~\TeV$). In Figs.~\ref{fig:bound-big} and \ref{fig:bounds} we present the bound from $B$ decays for $M_{H^+}=1200~\GeV$.

\subsection{Neutrino oscillations}

One of the most stringent bounds comes, perhaps
surprisingly, from neutrino oscillation experiments.  The new interaction will
change the neutrino matter potential which modifies the neutrino oscillation
pattern. It is useful to
express the new interaction in terms of the usual non-standard
interaction (NSI) operators which normalize the strength of the new matter potential to that induced by weak interactions. We define the NSI parameter by the
operator
\begin{equation}
\mathcal{L}_{\rm NSI} = 2\sqrt{2}G_F\varepsilon_{\alpha\alpha}^f\,\left(\bar\nu_{\alpha L}\gamma_\mu\nu_{\alpha L}\right)
  \left(\bar f \gamma^\mu f\right),
\end{equation}
and therefore we obtain
\begin{equation}\label{eq:epsilon}
  \varepsilon_{\alpha\alpha}^f=\frac{c_\alpha c_f}{g^2} \frac{4M_W^2}{M_X^2}.
\end{equation}
Due to the lack of flavor universality of the new gauge group we
expect a non-standard matter potential  (we remind the reader that
a universal diagonal matter potential has no impact on neutrino
oscillations)
\begin{equation}
  V_X \propto  \text{diag}
  \left(0,\quad 0,\quad \varepsilon_{\tau\tau}\right).
\end{equation}
It is important to mention that, as normal matter is neutral, the
kinetic mixing parameter $\varepsilon$ does not play any role in
neutrino oscillations. If we assume the number density of protons,
neutrons and electrons all to be the same, and use
Eqs.~(\ref{eq:Xcoupling}) and (\ref{eq:epsilon}), we can translate the non-universal matter
effects into the usual non-standard interaction parameter:
\begin{align}
  \varepsilon_{\tau\tau}&\equiv\varepsilon_{\tau\tau}^{p} +
  \varepsilon_{\tau\tau}^{n} + \varepsilon_{\tau\tau}^{e}\nonumber\\
  &=\frac{4M_W^2}{g^2 M_X^2} (-g_X)\left[c_{eR}+c_{eL}+
    3(c_{uR}+c_{uL}+c_{dR}+c_{dL})\right]
  = 3\frac{v_1^2 v^2}{v_1^2 v_2^2 + v_s^2 v^2}.
\end{align}
Atmospheric neutrinos play a major role in constraining the $\tau\tau$
NSI, leading to~\cite{Gonzalez-Garcia:2013usa}
\begin{equation}\label{eq:neutrino-bound}
  |\varepsilon_{\tau\tau}| < 0.09.
\end{equation}
Notice that the new matter potential does not depend on the gauge
coupling, but only on the VEVs of the scalar fields, analogous to what
happens with the standard matter potential. Note also that in the absence of the singlet scalar $s$,
the non-standard interaction would be $\varepsilon_{\tau\tau}=3v^2/v_1^2> 3$, which violates the
experimental limit of Eq.~(\ref{eq:neutrino-bound}), for any $M_X$. Plugging in numbers we
find $v_s>1.3(0.6)~\TeV$ for $\tan\beta=0.5(2)$.

\subsection{Electroweak $T$ parameter}

From the mass matrix (\ref{eq:mass-gauge}), we can see that the $Z$
boson mass is shifted from its SM value by
\begin{equation}
  \Delta M_Z^2 \simeq\frac{g_X^2}{9}\frac{v_1^4}{v^2},
\end{equation}
which contributes to the electroweak $T$ parameter. Therefore, the current
bound~\cite{Agashe:2014kda}~\footnote{$X$ is not expected to contribute to the running of electroweak parameters at low scales due to small $g_X$.}
\begin{equation}
T \simeq \frac{1}{\alpha}\frac{\Delta M_Z^2}{M_Z^2}= 0.01\pm 0.12
\label{eq:T}
\end{equation}
imposes a constraint $g_X<0.035$ for $\tan\beta=1/2$, with the
constraint becoming weaker for larger values of $\tan\beta =v_2/v_1$
as the fourth power.

\begin{figure}[!h]
 \begin{center}\vspace{-5mm}
   \includegraphics[scale=0.65]{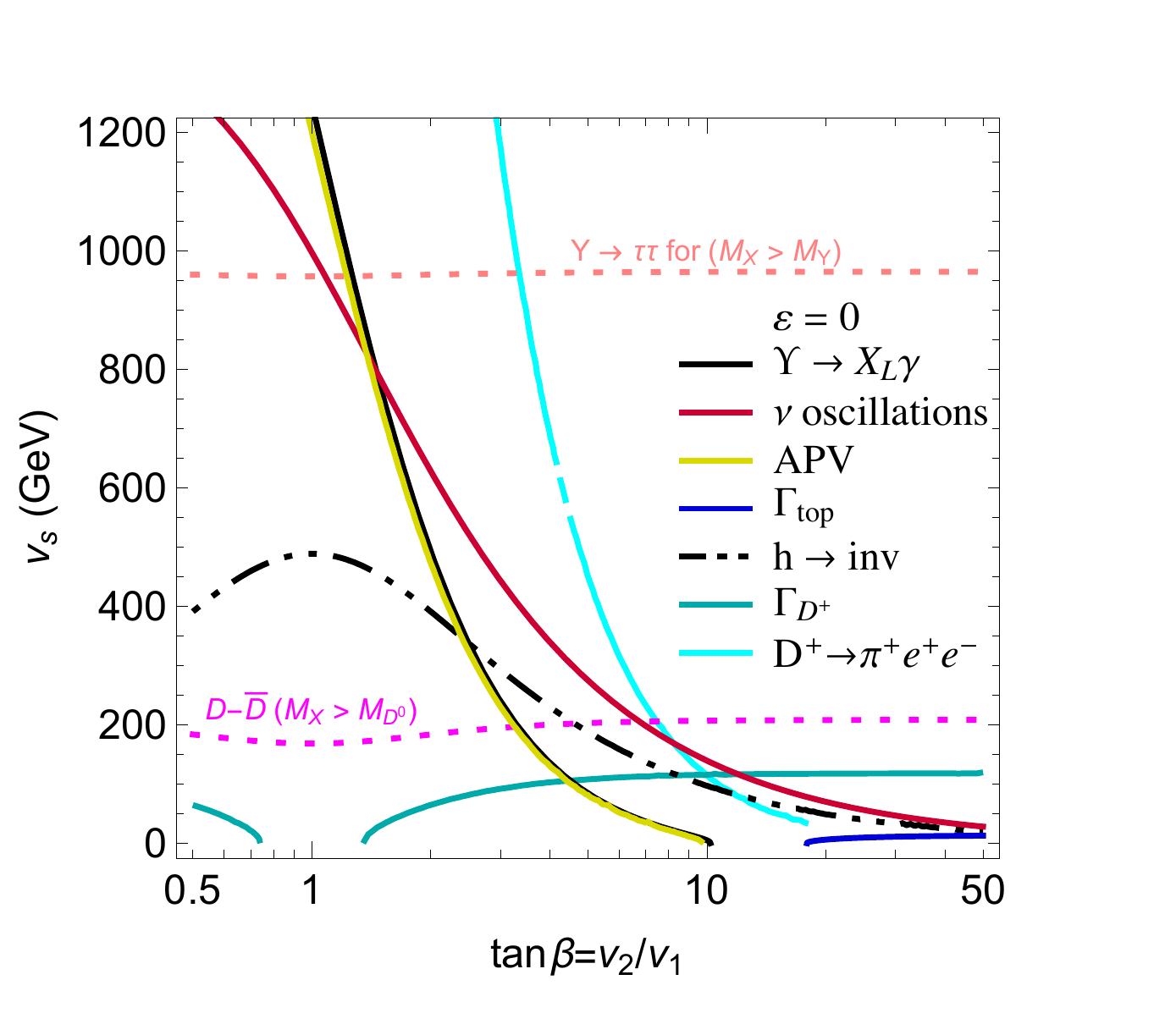}
 \end{center} \vspace{-0.5cm}
 \caption{Constraints on the VEV of the SM singlet scalar $s$ as a
   function of $\tan\beta\equiv v_2/v_1$ from $\Upsilon\to X_L\gamma$
   decay, atomic parity violation (APV), $D^+\to\pi^+\mu^+\mu^-$, total $D^+$ width,
   neutrino oscillations, top quark total width, and Higgs invisible decays (see Sec.~\ref{sec:appendix}). We also show
   the bounds on $v_s$ from
   $\Upsilon\to\tau^+\tau^-$ and $D^0-\overline{D^0}$ mixing, although
   they only apply for $M_X\gg M_\Upsilon$ and $M_X\gg M_{D^0}$,
   respectively (see Fig.~\ref{fig:bound-big} and \ref{fig:bounds}).}
 \label{fig:tanb}
\end{figure}

\vspace{1cm}

The constraints derived here are plotted
in the $g_X-M_X$ plane in Fig.~\ref{fig:bound-big} and \ref{fig:bounds}.  The origin of various constraints are
labeled. The four panels correspond to four values of $\tan\beta =
v_2/v_1$.  The label on the right indicates the $Z-X$ mixing angle
$s_X$.  Note that some regions in this plane are excluded
theoretically, since $M_X$
must obey an inequality.

We present in Fig.~\ref{fig:tanb} the constraints from $\Upsilon$, $D^+$, top and Higgs decays (see Sec.~\ref{sec:appendix}), atomic parity violation, neutrino oscillations and
$D^0-\overline{D^0}$ mixing on $v_s$ as a function of $\tan\beta$.
We do not show the $K^+\to\pi^+X$ bound, but we notice that it is much stronger than the others for $M_X<114~\MeV$ and $151<M_X<260~\MeV$. Also, the $B^+\to K^+ X$ is omitted  due to the strong dependence with $M_{H^+}$.
Outside the Kaon bound region, we see clearly that $D^+\to\pi^+e^+e^-$ dominate for $\tan\beta<8$, as long as $M_X<m_{D^+}-m_{\pi^+}$. The total $D^+$ width dominates for $\tan\beta>13$ if $M_X<m_{D^+}-m_{\pi^+}$.
The neutrino oscillations bound is independent of $M_X$. It is the main constraint for $8<\tan\beta<13$, or $1.5<\tan\beta$ if the $D^+$ channels are forbidden. The region $\tan\beta<1.5$ is well covered for any $M_X$ by the combination of  APV and $\Upsilon\to X\gamma$. Higgs invisible branching ratio and top width provide complementary constraints for large $\tan\beta$. Once the  $X$ boson mass exceed $M_{D^0}$ or $M_\Upsilon$, $D^0-\overline{D^0}$ mixing and $\Upsilon \to \tau \tau$ dominate the constraints for $\tan\beta$ above $7.5$ and $3.2$, respectively.

\section{Other constraints}
\label{sec:appendix}

\begin{figure}[!t]
 \begin{center}
   \includegraphics[scale=0.65]{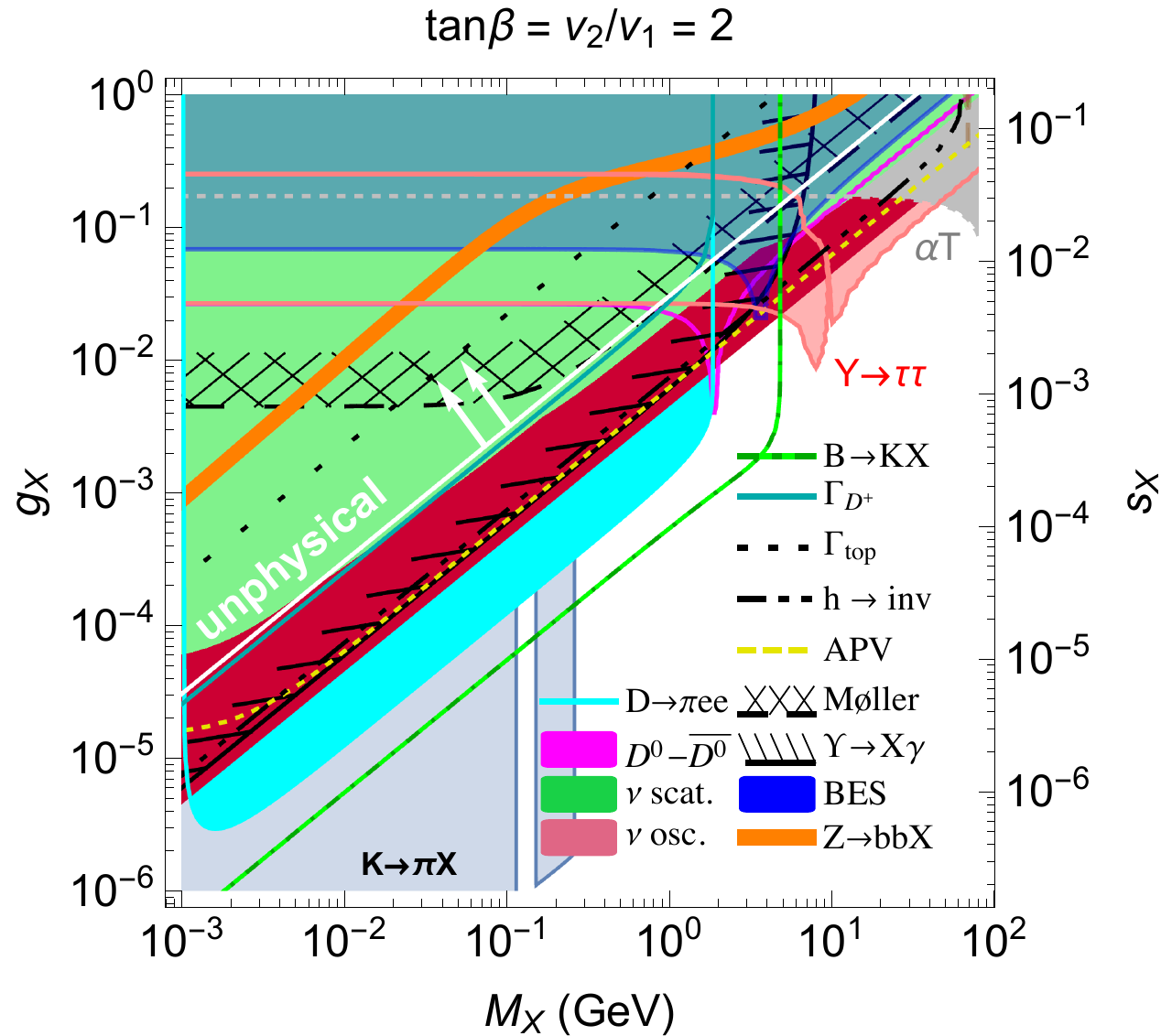}\includegraphics[scale=0.65]{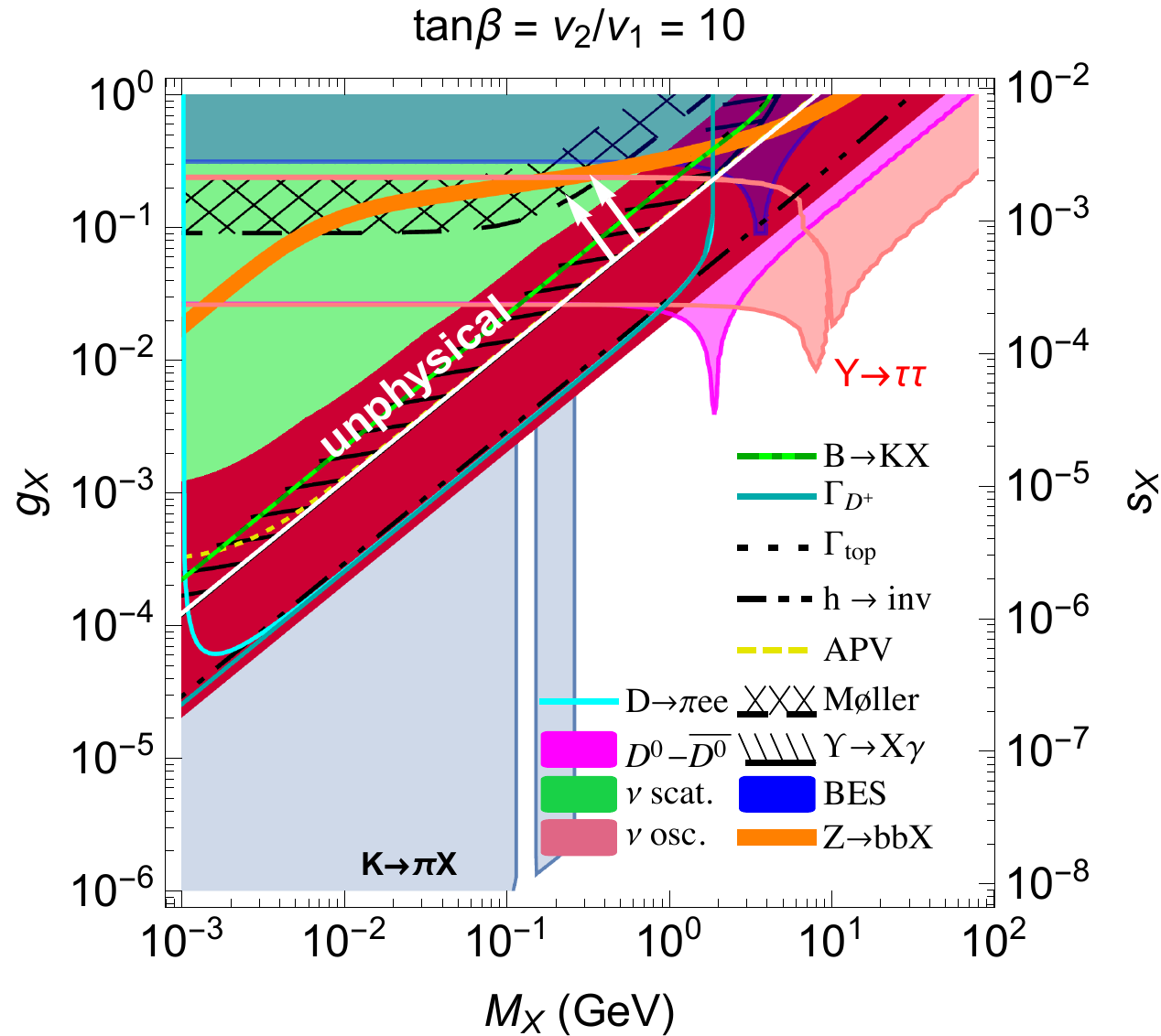}
 \end{center} \vspace{-0.7cm}
 \caption{More complete list of constraints on the $U(1)_{B-L}^{(3)}$ gauge boson mass
   $M_X$ and coupling $g_X$ for $\tan\beta=2,10$. For
   convenience the $X-Z$ mixing, $s_X$, is also shown.  Notice that
   for a given $g_X$, the mass of the gauge boson $M_X$ is bounded
   from below, so there is an unphysical region in the upper left
   corner of the $M_X\times g_X$ plane (delineated by the white
   line). The ``$\nu$ scat'' bound is a combination of all neutrino
   scattering experiments listed in the text, while ``$\nu$ osc.''
   comes from non standard interaction effects (matter potential) on
   atmospheric neutrinos. ``APV'' refers to atomic parity violation.
   Here the charged Higgs mass, relevant to the $B\to K X$ constraint, is taken to be 1200~GeV.}
 \label{fig:bounds-busy}
\end{figure}

Here we provide a more complete analysis, including those constraints which turned out
{\it a posteriori} to be not as stringent as the ones discussed in the previous section.
In Fig.~\ref{fig:bounds-busy} we present the bounds from the previous section together with a few bounds from this section (chosen by their importance in ruling out the physical region of the parameter space).

\subsection{Atomic parity violation}

An important process is atomic parity violation (APV), in which
the weak charge, especially for $^{133}$Cs, has been measured very
precisely. The standard model prediction is $Q_W^{SM}=-73.16\pm0.3$,
while the experimental measurement combined with theoretical
calculations yield $Q_W=-73.16\pm0.35$~\cite{Porsev:2010de}. In
our model, since the $X$ boson mixes with the $Z$, there are new
contributions to $Q_W$. The fractional
contribution of the $X$ mediated APV is given by
\begin{equation}
  f_{APV} = 1+s_X^2\frac{M_Z^2}{M_X^2+q^2}=1\pm0.0063,
\end{equation}
where $\vev{q^2}\simeq (2.4~\MeV)^2$ is the estimated average squared
momentum transfer. This allows us to put a direct bound
$v_s>2(0.5)$~TeV at 90\% CL for $\tan\beta=0.5(2)$ and for
mediator squared-masses above $\vev{q^2}$.

\subsection{Flavor changing top decay}

The $X$ boson can also mediate flavor-changing processes involving the top quark. The decay
$t \rightarrow c X$ is predicted in the model. The width for this decay can be calculated directly, or using the equivalence theorem and the Goldstone boson coupling Eq.~(\ref{eq:goldstone-fermions}),
\begin{equation}
\Gamma(t \rightarrow c X) \simeq \frac{g_X^2}{288 \pi} \frac{|V_{cb}|^2 m_t^3}{M_X^2}.
\end{equation}
For $g_X = 10^{-3}$ and $M_X = 100$ MeV, the width is 0.9~MeV,
corresponding to a branching ratio of $6.5 \times 10^{-4}$, which
would not be easy to observe.  However, if the mass of $X$ is lower,
this branching ratio increases.  For example, when $M_X = 1$ MeV, top
quark width would set a constraint on $g_X$ to be less than about
$2\times10^{-4}$.  The new contribution to the top quark width cannot
exceed 0.38 GeV (at 2 sigma) \cite{Agashe:2014kda}.
The top width provides a direct bound
  $v_s$, which turns out to be important only for large values of
  $\tan\beta$ (see Fig. \ref{fig:tanb}). Note that this decay can be
understood in terms of Goldstone boson equivalence theorem, as the top
decays primarily into the longitudinal $X$. Apart from $t\to cX$, as the Higgs has flavor changing couplings (see Eq.~(\ref{eq:higgs-eft})), $t\to ch$ transitions are also possible. Nevertheless, the flavor changing Yukawa is doubly suppressed, by $V_{cb}$ and by the small mixing angle between $H$ and $H'$, making this branching ratio typically small, below $10^ {-4}$.

\subsection{$h\to X X$ decay}
The presence of $X-Z$ mixing will lead to Higgs decays to $X$ pairs (dominantly to the longitudinal modes). The $X$ bosons typically further decay to neutrinos, thus leading to a contribution to the invisible Higgs branching ratio which is bounded to be smaller than 0.28~\cite{Aad:2015txa}. The invisible width is given by (see Eq.~\ref{eq:h-goldstone})
\begin{equation}
  \Gamma(h\to XX)=\frac{g_X^4}{2592 \pi}\frac{v_1^4 v_2^4}{v^6 M_h}\left(\frac{M_h^4-4M_X^2 M_h^2+12 M_X^4}{M_X^4}\right)\sqrt{1-\frac{4M_X^2}{M_h^2}}.
\end{equation}
In the limit of $M_X\ll M_h$ this becomes
\begin{equation}
  \Gamma(h\to XX)=\frac{M_h^3\sin^4(2\beta)}{32\pi v^2[\sin^2(2\beta) +4v_s^2/v^2]^2},
\end{equation}
which translates to $v_s > \sin(2\beta)\times490~\GeV$. Notice that the Higgs can also decay to
$XZ$ via the mixing with ${\rm Re}(s)$, leading to interesting modifications of Higgs
phenomenology (e.g. $h\to XZ\to \tau^+\tau^- \ell\ell$, with the $\tau$ pair invariant mass at $M_X^2$).
Nonetheless, since this mixing is a free parameter, we do not consider this channel.

\subsection{M{\o}ller scattering}

Measurements from SLAC
E158~\cite{Anthony:2003ub} are sensitive to modifications of the
parity-violating asymmetry in low scale $e^- e^-$ scattering,
\begin{equation}
  A_{PV}=\frac{\sigma_R-\sigma_L}{\sigma_R+\sigma_L}=(-175\pm30_{\rm stat}\pm20_{\rm syst})\times 10^{-9},
\end{equation}
where $\sigma_{R,L}$ indicate the cross section for incident right-
and left-handed electrons. The asymmetry is dominated by the
interference term between the photon and the $Z$. In this
experiment, the sensitivity to $s_w^2$ is significantly enhanced due to an accidental
cancelation in the factor $(1/4-s^2_w)$ appearing in the
asymmetry. In our model, this cancelation plays no role in the
sensitivity to the $X$ boson contributions, as the parity-violating
coupling comes entirely from the mixing with the $Z$. Because of that,
the $X$ fractional contribution to $A_{PV}$ is basically the mixing
with the $Z$ and the ratio of propagators,
\begin{equation}
  \frac{A_{PV}^{\rm SM+X}}{A_{PV}^{\rm SM}}-1 = \frac{s_X^2M_Z^2}{q^2+M_X^2}<0.21,
\end{equation}
where we added the statistical and systematical fractional errors in
quadrature. The average momentum transfer is $\langle q^2\rangle =
(0.161~\GeV)^2$.

\subsection{$Z$ decays to $\tau^+\tau^- X$ and $b\bar bX$}

For $M_X$ below the $Z$ mass, the
processes $Z\to \tau^+\tau^- X$ and $Z\to b\bar bX$ may also constrain our model.
When $M_X\ll M_Z$, these processes will measure the diagonal Yukawas between the third family fermions and $G_X$, the Goldstone mode of $X$, see Eq.~\ref{eq:goldstone-fermions}.

In this limit, the partial widths above can be written as~\footnote{The log divergence should be regulated by the 1-loop amplitude. We estimate the effect to be small for the range of parameters chosen here.}
\begin{equation}
  \Gamma(Z\to f\bar fX)=\frac{N_c}{192\pi^3}M_Z |y_f^{G_X}|^2
  \left[g_V^{f\,2}\left(1+\log\frac{M_Z^2}{M_X^2}\right)+g_A^{f\,2}\left(-\frac{14}{3}+\log\frac{M_Z^2}{M_X^2}\right)\right],
\end{equation}
with
\begin{equation}
  g_V^\tau = \frac{g}{4c_w}(4s_w^2-1), \quad g_A^\tau = \frac{g}{4c_w}, \quad
  g_V^b = \frac{g}{4c_w}\left(\frac{4s_w^2}{3}-1\right), \quad g_A^b = \frac{g}{4c_w}.
\end{equation}

In the limit of heavy $X$, the Goldstone modes contribute very little, and the mass of the fermions can be safely neglected. In this case, the results of Ref.~\cite{Marciano:1979cd} on $Z\to W\ell\nu$
can be easily recast into our scenario leading to
\begin{equation}
  \frac{d\Gamma(Z\to f\bar f X)}{dx}=\frac{M_Z}{6\pi^3}
            \left[(g_V^f c_V^f+g_A^f c_A^f)^2(h_1(x) + h_3(x))\right],
\end{equation}
where $x=2E_X/M_Z$ is the energy fraction carried by $X$, and
\begin{equation}
  c_V^f\equiv(c_{fR}+c_{fL}),\quad c_A^f\equiv(c_{fR}-c_{fL}),
\end{equation}
where $c_\alpha$ is defined in Eq.~(\ref{eq:Xcoupling}). The
 functions $h_1(x)$ and $h_3(x)$ are given in
Eqs. (10.1), and (10.3) of Ref.~\cite{Marciano:1979cd}.
The total width is obtained by
integrating the differential width in $x$ from $2M_X/M_Z$ to
$1+M_X^2/M_Z^2$.

There are no dedicated searches for these channels. For the $Z\to\tau^+\tau^-$,
we require the additional width not to exceed the experimental
uncertainties of $0.2$~MeV. In the case of $Z\to b\bar b$, the uncertainty on $R_b$ imposes the
additional width to be below $2.8$~MeV. In Fig.~\ref{fig:bounds} we show only the constraint from
$Z\to b\bar bX$, as it is slightly more stringent than $Z\to \tau^+\tau^- X$. These effects could be of
particular interest to the models discussed in Refs.~\cite{Farzan:2015doa, Farzan:2015hkd}.
It would also be interesting
to search for such a light gauge boson in the decay of $Z$, since the new
decay mode can be distinguished from the two-body decay mode $Z\rightarrow f\bar f$
by its distinct kinematic shape.\footnote{We thank A. Khanov for discussion on this prospects.}

\begin{figure}[t]
 \begin{center}
   \includegraphics[scale=0.35]{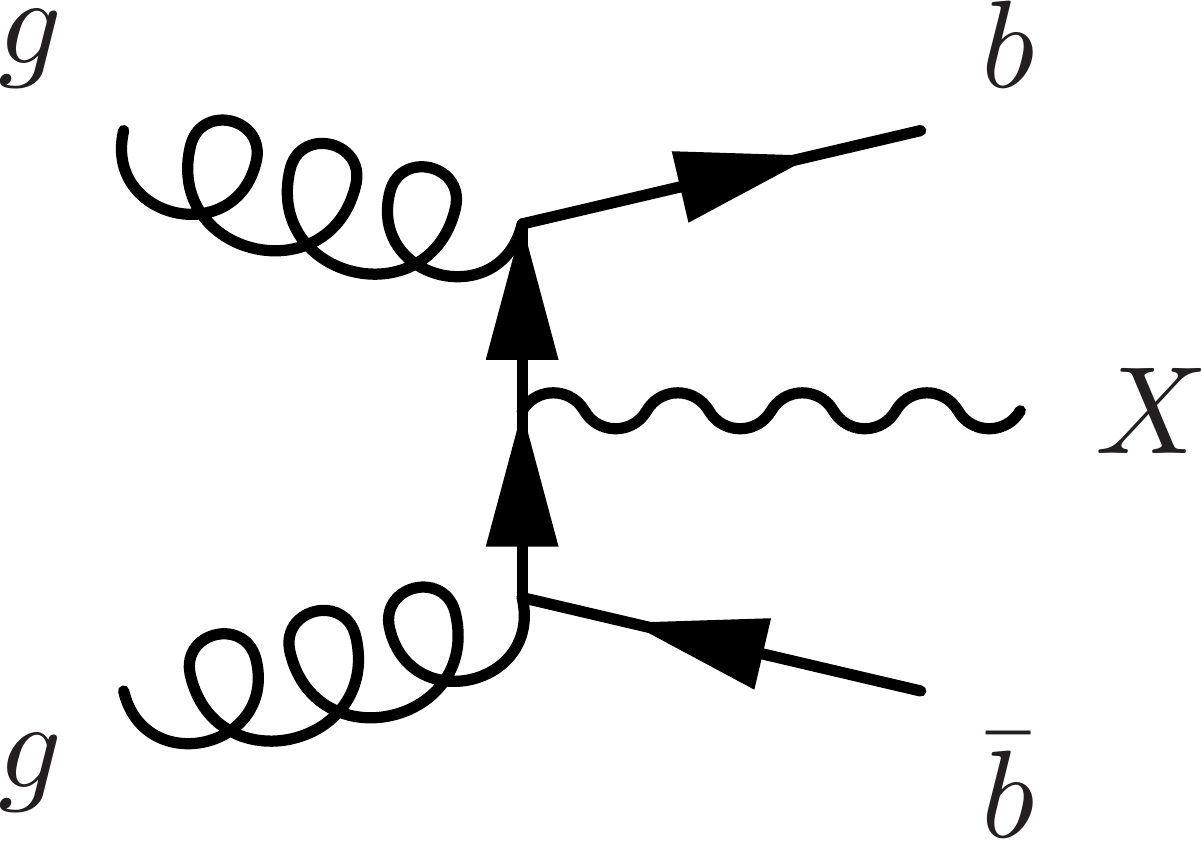}
 \end{center}
 \caption{Dominant $X$ production mode at the LHC for $m_X$ at the TeV
   scale.}
 \label{fig:feyn-diag}
\end{figure}

\subsection{$X$ resonant production at the LHC}

If the $X$ boson is above the electroweak scale, the LHC will
eventually provide the best bound on the direct production of $X$. The
$T$ parameter imposes the mixing with the $Z$ to be small, and
therefore the couplings of a heavy $X$-boson to the third family
fermions will dominate the phenomenology. Although a comprehensive LHC
analysis is beyond the scope of this paper, we point out that a search
for a resonance decaying to $\tau^+\tau^-$ in association with two
$b$-jets seems to be a promising way of exploring this model, see
Fig.~\ref{fig:feyn-diag}. In fact, the $b\,\bar b\,\tau^+\tau^-$
final state has been studied by CMS~\cite{Chatrchyan:2012sv} and
ATLAS~\cite{ATLAS:2013oea} in third generation leptoquark searches,
but since each $b\,\tau$ pair reconstructs a resonance it is not
straightforward to interpret these results in the context of our
model.

Should the $X$ gauge boson have already shown up in the dimuon searches at the LHC?  At 13 TeV LHC, the cross section for $b \bar{b}$ production is about 154 micro-barn.  The $X$ boson may be emitted from the $b$ quark.  We estimate the production cross section for $p p \rightarrow b \bar{b} X$ to be $\sim 154 \mu{\rm b} \times g_X^2/(36 \pi)$.  For $g_X = 0.05$, number of $X$ bosons produced with 40 fb$^{-1}$ of data is about $10^8$.  The branching ratio for $X \rightarrow \mu\mu$ is about $10^{-3}$ ($10^{-6}$) for $\tan\beta = 2 (10)$, which would imply that the number of dimuon events is  about $10^5$ ($10^2$).  The background for dimuon resonance searches is a few times $10^5$ events per GeV at low mass, and thus the $X$ boson would not have been observed. With more data, for a range of model parameters, the $X$ boson may be observable at the LHC as a dimuon resonance.

\subsection{Meson-antimeson oscillations}

The presence of FCNC in scalar and gauge
boson interactions can modify $K^0-\overline{K^0}$, $B_d-\overline{B}_d$, $B_s-\overline{B}_s$, and $D^0-\overline{D^0}$
oscillations.  The case of $D^0-\overline{D^0}$ mixing, which provides the best limits on the model, is already analyzed
in Sec. \ref{sec:pheno}.  Here we complete this analysis.

The general scalar contributions to meson-antimeson mixing is given in Eq. (\ref{eq:meson}).
 The vector boson $X$ will also contribute to the meson
oscillation via $s$-channel exchange ~\cite{Nir:1990yq}
\begin{equation}
    (\Delta m_S)_X=\frac{\sqrt{2}}{6}G_F f_S^2 m_S B_S \eta_S
  \frac{M_Z^2}{m_S^2-M_X^2}\left| \frac{2g_X^2 U^X_{ij}/3}{g/c_w}\right|^2,
\end{equation}
where $U^X=V^L_{u,d}.{\rm diag}(0,0,1).V^{L\dagger}_{u,d}$. In fact,
this contribution is suppressed by both the small mixing, $U^X_{ij}$
and $g_X$.  Except for the case of $D^0-\overline{D^0}$ mixing where the
$X$ boson exchange becomes important for $M_X \sim$ MeV, this contribution
is generally sub-leading.

Using the parameters found in refs.~\cite{Babu:2009nn, Babu:1999me}
and imposing that the extra contribution is smaller than the
experimental and theoretical uncertainties~\cite{Buras:1990fn,
  Lenz:2012mb} we find that
\begin{align}
  K-\bar{K}: \quad & \left(\frac{100\,\GeV}{m_\varphi}\right)
        {\rm Re}\left(\frac{h_{21}^d}{\sqrt{2}m_s/v}\right)\lesssim 1.4\times 10^{-2}\\
  B_d-\bar{B}_d: \quad  & \left(\frac{100\,\GeV}{m_\varphi}\right)
        {\rm Re}\left(\frac{h_{31}^d}{\sqrt{2}m_b/v}\right)\lesssim 3.1\times 10^{-3}\\
  B_s-\bar{B}_s: \quad& \left(\frac{100\,\GeV}{m_\varphi}\right)
        {\rm Re}\left(\frac{h_{32}^d}{\sqrt{2}m_b/v}\right)\lesssim 1.3\times 10^{-2}\\
  D-\bar{D}: \quad&\left(\frac{100\,\GeV}{m_\varphi}\right)
        {\rm Re}\left(\frac{h_{12}^u}{\sqrt{2}m_c/v}\right)\lesssim 3.4\times 10^{-3}
\end{align}
In our benchmark points, all contributions to down flavored meson oscillation
vanish, since $V_{ub}$ and $V_{cb}$ are generated in the up-quark
sector.

\subsection{Tau physics}

 Precise measurements of the $\tau$ mass and
production cross section were performed by the BESIII
collaboration~\cite{Ablikim:2014uzh}. Doing a scan in the energy of
the $e^+e^-$ beam around the $\tau$ threshold made it possible to measure
the $\tau\tau$ production cross section at the sub-percent level. To
estimate the constraint from BESIII, we require the ratio of BSM
and standard cross sections $\sigma(e^+e^-\to
A,X,Z\to\tau^+\tau^-)/\sigma_{SM}(e^+e^-\to A,Z\to\tau^+\tau^-)$ not
to exceed the experimental errors at fixed $\sqrt{s}$, namely 3.1039,
3.542, 3.553, and 3.5611~GeV.

\subsection{$(g-2)_\mu$}
The $X$ boson may contribute to the muon anomalous magnetic moment through its mass mixing with
the $Z$. In contrast to the case of pure vectorial couplings, the axial-vector contribution to $(g-2)_\mu$
does not saturate for small $M_X$~\cite{Studenikin:1998cs, Jegerlehner:2009ry}, growing as $M_X$
diminish. For $M_X\ll m_\mu$, requiring the modification to $a_\mu$ not to exceed $10.8\times 10^{-8}$
the constraint is approximately
\begin{equation}
  g_X<1.2\times 10^{-4} (1+\tan^2\beta)\left(\frac{M_X}{\rm MeV}\right)~~~~~~{\rm from}~(g-2)_\mu.
\end{equation}
Nevertheless, for a given $M_X$ the $g_X$ coupling is bounded from above by (see eq.~(\ref{eq:mx}))
\begin{equation}
  g_X<1.2\times 10^{-5} \left(\frac{1+\tan^2\beta}{\tan\beta}\right)\left(\frac{M_X}{\rm MeV}.\right)~~~~~~{\rm from~eq.}~(\ref{eq:mx}),
\end{equation}
Thus, for any reasonable value of $\tan\beta$, the bound from the muon anomalous magnetic moment is
always in the ``unphysical'' region of Fig.~\ref{fig:tanb}.

\subsection{Neutrino--electron scattering}

The neutrino electron scattering
cross section may be considerably modified in the presence of the
extra gauge boson $X_\mu$.
We have estimated the constraint coming from a given experiment by
making the simplified assumption of a fixed momentum transfer. Due to
the large number of experiments, we only state the numbers we
use. All limits can be found in Fig.~\ref{fig:bounds}.

The limits from solar neutrino measurements at the Borexino
experiment~\cite{Bellini:2011rx} were calculated in
Ref.~\cite{Harnik:2012ni} for the universal $B-L$ scenario. In our
case, we estimated it by requiring that the $\nu-e$ scattering cross
section does not exceed $10\%$ of the standard cross section for
$q^2=2 m_e E_{\rm rec}$, where $m_e$ and $E_{\rm rec}\sim 300~{\rm
  keV}$ are the electron mass and recoil energy. Limits coming from
reactor neutrinos at the GEMMA experiments~\cite{Beda:2009kx} were
also calculated for a universal $B-L$ light gauge boson in
Ref.~\cite{Harnik:2012ni}. We have checked that in the $\MeV-\GeV$
region, the GEMMA experiment is always less sensitive than Borexino.

\begin{table}[!h]
  \begin{center}
    \begin{tabular}{| C{4cm} | L{11.8cm} |} \hline
Experiment & Constraint \\ \hline \hline
 M{\o}ller scattering & $Z-X$ mixing leads to parity violation in $e^-e^-$ scattering \\ \hline
 $t\to cX$ & Flavor changing $c\,t X$ coupling can contribute to the total top width, which is bounded as $\Delta \Gamma_t<0.44$~GeV~\cite{Agashe:2014kda}\\ \hline
 $Z\to f\bar f X$ & There is no dedicated search for  $Z\to\tau^+\tau^-+/\!\!\!\!E_T$ ($Z\to b\bar b+/\!\!\!\!E_T$).
A direct bound on $g_X$ may be obtained by requiring these branching ratios
not to exceed $0.2~\MeV$ $(2.8~\MeV)$.\\\hline
 $h\to XX$ & Decays to longitudinal $X$ pair contributes to invisible width.\\\hline
$X$ at the LHC & Resonant production of $X$ decaying to $\tau^+\tau^-$ in association with two $b$-jets at the LHC may constrain the parameter space for realizations of the model at the TeV scale \\ \hline
Fixed target & Much weaker than in kinetic mixing scenario~\cite{Bjorken:2009mm} as
${\rm BR}(X\to\nu\nu)$ typically dominates, especially for large $\tan\beta$. \\\hline
$(g-2)_e$ and $(g-2)_\mu$ & ~The axial-vector contribution does not saturate for small $M_X$~\cite{Studenikin:1998cs, Jegerlehner:2009ry}, but the bound is nevertheless weak.\\\hline
BESIII & $e^+e^-\to\tau^+\tau^-$ near the $\tau$ threshold~\cite{Ablikim:2014uzh}.\\ \hline
$K,B_d,B_s$ oscillation & May lead to strong bounds on off-diagonal Yukawa couplings, forcing $V_{CKM}$ to be generated in the up sector. The contribution from heavy scalar exchange is both loop and CKM suppressed.\\
\hline
Neutrino scattering & Borexino~\cite{Bellini:2011rx, Harnik:2012ni},
GEMMA~\cite{Beda:2009kx}, CHARM~II~\cite{Vilain:1994qy,
  Davidson:2003ha}, TEXONO~\cite{Deniz:2009mu},
MiniBooNe~\cite{AguilarArevalo:2007it, Aguilar-Arevalo:2013pmq}, and
LSND~\cite{Auerbach:2001wg} constrain $\nu-e$ scattering. NUTEV data
displays a $2.7\sigma$ tension with the SM
prediction~\cite{Zeller:2001hh}. We require this tension not to be
worsened by $1.6\sigma$.\\\hline
$W\to \tau\nu X$ & For $M_X\ll M_W$ probes the Yukawa coupling to $\tau$, but does not constrain the model significantly.\\\hline
$t\to bWX$ & For $M_X\ll m_t$ probes the Yukawa coupling to the top, but does not constrain the model significantly.\\\hline
LEP & LEP bound on resonant $e^+e^-\to\tau^+\tau^-$ production~\cite{Acciarri:2000uh}
 can be used to put bounds on the $X$ coupling to leptons.\\\hline
$\pi\to X\gamma$ & The bound derived in the case of pure kinetic mixing~\cite{Batell:2009di} can be easily translated to our scenario, leading to a very weak bound in the whole parameter space. \\\hline
    \end{tabular}
  \end{center}	
  \caption{List of constraints on the $U(1)_{B-L}^{(3)}$ gauge and
    scalar sector. For the neutrino scattering bounds, the
    approximation of mean momentum transfer $\langle q^2\rangle$ was
    made separately for each experiment. See
    Sec.~\ref{sec:appendix} for details.}
\label{tab:bounds}
\end{table}

The bound from CHARM~II~\cite{Vilain:1994qy} was computed for the case
of NSI in Ref.~\cite{Davidson:2003ha}, yielding
\begin{equation}
  -0.025 < \varepsilon_{\mu\mu}^{eL}<0.03, \qquad -0.027<\varepsilon_{\mu\mu}^{eR}<0.03.
\end{equation}
This can be easily translated to our case using Eq.~(\ref{eq:epsilon})
and fixing $q^2=0.01~\GeV^2$.

The TEXONO experiment~\cite{Deniz:2009mu} measured $\bar\nu_e$
scattering on electrons in a CsI detector with $q^2 \approx
3~\MeV^2$. The ratio of the experimental cross section to the SM one
was found to be
\begin{equation}
  \frac{\sigma_{\rm exp}}{\sigma_{SM}}=1.08\pm0.21({\rm stat})\pm0.16({\rm syst}).
\end{equation}
There are two other measurements of the TEXONO experiment using a high
purity Ge detector~\cite{Wong:2006nx} and a N-type point-contact Ge
detector~\cite{Chen:2014dsa}. We have checked that these bounds are
negligible for $X$ at or above the MeV scale.

\clearpage

The MiniBooNe experiment~\cite{AguilarArevalo:2007it,
  Aguilar-Arevalo:2013pmq} has measured a variety of neutrino cross
sections, ranging from $\nu_\mu$ neutral current scattering on nucleus
to $\nu_e-e$ elastic scattering. Due to the mass dependence,
scattering on electrons will have a lower $q^2$ for the same recoil
energy and thus will be more important for a lighter mediator. We
assumed conservatively $q^2 = 2m_eE_{th}$, where $E_{th}= 140~\MeV$ is
the experimental threshold energy of the scattered electron, and a 10\%
error on the elastic cross section.

The NUTEV experiment measured the ratio of neutral to charged current
cross sections for $\nu$ and $\bar\nu$ to $1\%$ precision and with a
mean $\langle q^2 \rangle\approx -20~\GeV^2$~\cite{Zeller:2001hh}.  In
fact, the NUTEV measurement of $(g_L^{\rm eff})^2$ displays a tension
with the standard model prediction at the $2.7\sigma$ level  (see
Ref.~\cite{Zeller:2001hh} for details).  A positive $g_X$ enhances
this cross section making the tension worse. We use the NSI bound from
Ref.~\cite{Davidson:2003ha}, namely,
\begin{equation}
  |\varepsilon_{\mu\mu}^q|<0.003,\quad q=u,d.
\end{equation}

Finally, the measurement of the $\nu-e$ elastic scattering cross
section, with a $\sim 17\%$ precision, by LSND~\cite{Auerbach:2001wg}
can also be used to constrain non-standard interactions in the
neutrino sector~\cite{Davidson:2003ha}. In the light mediator
scenario, the LSND bound is somewhat special due to its low threshold
for the electron recoil energy, $E_{th}=18~\MeV$. The limits found in
Ref.~\cite{Davidson:2003ha} can be applied to our scenario by using
Eq.~(\ref{eq:epsilon}) with $q^2=2 m_e E_{th}$.

\subsection{$t\to bWX$}

Another 3-body decay that can be enhanced by the Goldstone coupling is $t\to bWX$. The dominant contribution comes from the $X$ emission by the initial top quark. The differential width is given by
\begin{equation}
  \frac{d\Gamma(t\to bWX)}{dxdy}=\frac{g^2}{128\pi^3}m_t\left(\frac{g_X}{3}\frac{m_t}{M_X}\frac{v_1^2}{v^2}\right)^2\frac{J}{(2-2(x+y)^2-r_X)^2+r_\Gamma},
\end{equation}
where $x\equiv E_W/m_t$ and $y\equiv E_b/m_t$ are the energy fractions carried by the $W$ and the $b$ quark,
\begin{equation}
  J= (4x-1)(x+y-1)+r_W(2-x-3y) + \frac{1}{r_W}(2y-1)\left[2x^2+x(4y-3)+2y^2-3y+1\right],
\end{equation}
and we define $r_X\equiv M_X^2/m_t^2$, $r_W\equiv M_W^2/m_t^2$, and $r_\Gamma=\Gamma_t^2/m_t^2$, with $\Gamma_t$ being the top width of $1.41~\GeV$. The mass of the $b$ quark was neglected. To obtain the total width, the integration should be performed within
\begin{align}
  \sqrt{r_W}\le \,&x\le (1+r_W)/2,\\
  (1-x-\sqrt{x^2-r_W})/2\le \,&y\le (1-x+\sqrt{x^2-r_W})/2.
\end{align}
For $M_X=1~\MeV$ and $\tan\beta=1$, requiring the additional width not to exceed the total top width uncertainty of $0.38~\GeV$ yields $g_X<1.1\times 10^{-3}$.

\subsection{$W\to\tau\nu X$}

Similar to the $Z\to f\bar f X$ decay, the $W$ can decay to $\tau \nu X$, where the longitudinal $X$ dominates for low masses.  The width is given by
\begin{equation}
  \Gamma(W\to\tau\nu_\tau X)=\frac{N_c\, g^2}{1536\pi^3}M_W\left(\frac{g_X}{3}\frac{m_\tau}{M_X}\frac{v_1^2}{v^2}\right)^2\left(-\frac{11}{6}+\log\frac{M_W^2}{M_X^2}\right).
\end{equation}
Notice that the coupling proportional to $B-L$ does not have the longitudinal enhancement. The decay proceeds through $Z-X$ mixing.
The $X$ typically decays to neutrinos, so the signature would still be $\tau+/\!\!\!\!E_T$, but with a distinct $p^\tau_T$ distribution. Since there is no dedicated search for such signature, we demand the width not to exceed $20~\MeV$, the uncertainty on the $W\to\tau\nu$ partial width. The bound is nevertheless weak: for $M_X=1~\MeV$ and $\tan\beta=2$ it leads to $g_X<0.01$.

\section{Outlook}
\label{sect:conclusions}

Inspired by the anomaly cancelation within a single standard model generation and the fact that
the third generation appears different with heavy masses and small mixings with the first two generations in the quark sector,
we have proposed and analyzed a gauged $U(1)_{B-L}^{(3)}$ symmetry that acts only on the third family.
We have constructed a class of fully consistent flavor models below the
weak scale,
which is renormalizable and exhibits a light gauge boson that couples non-universally to the quark and lepton flavors.

Our model exhibits a very rich phenomenology,  yielding many interesting observables of different types.
For instance, to accommodate the observed mixing of generations in the presence of the new flavor-dependent $U(1)$ gauge symmetry
it is necessary to extend the Higgs sector. The minimal extension involves  a second Higgs doublet (for the mixing) as well as an additional
SM singlet which is charged under this $U(1)$,  for consistent symmetry breaking and phenomenology.
While two Higgs doublet models, with or without additional singlets, have been extensively studied in the literature,
our realization has a number of unique features due to the unusual flavor structure.
The extended Higgs sector of this model can be studied at the LHC, as well as with precision electroweak data and flavor observables. Here, we only sketched the relevant bounds; an in-depth analysis will be desirable.

Another class of relevant observables that is sensitive to the structure of the extended Higgs sector is based on precision meson decay data. This includes $K^+\to\pi^+ X$, $B^+\to K^+ X$,  $t\to c X$, $\Upsilon\rightarrow X\gamma$, $D^+\to\pi^+ X$ decays,  among others.  For decay processes, at high energies, the
equivalence theorem implies that the longitudinal mode can be replaced by the Goldstone boson associated with the breaking of the symmetry. Thus, many of the constraints would survive even in the limit of a global symmetry.
When the decay channel $K^+\to\pi^+ X$ is kinematically viable, kaon physics poses by far the most
important bound on the model.  $B^+\to K^+ X$ is also quite stringent, but depends strongly on $M_{H^+}$ and $\tan\beta$.

When $X$ is heavier, for a range of parameters with $g_X \sim 10^{-3}-10^{-2}$ and the $X$ gauge boson
mass of order $300~\MeV-1~\GeV$, neutrino oscillations become the main probe even at present.
The induced neutrino non-standard interaction is flavor conserving, namely $\epsilon_{\tau\tau}$ in the present model, and its values are in the interesting range for
DUNE, Hyper-Kamiokande, PINGU,  and other present and future experiments with increased sensitivity.
While the gauge symmetry is necessary to induce the new matter effect, the observables only depend on the size of the scalar VEVs and not on the gauge coupling.
The neutrino NSI thus probe the scale of the symmetry breaking, which can be even above a few TeV and still lead to potentially observable effects.

As already mentioned in the Introduction, our model has an important ambiguity: which leptons should carry the new gauge quantum numbers. While for quarks the choice is clear -- the top and bottom have very small mixing with the quarks from the other generations, which we are trying to explain --
in the case of leptons there is no natural choice. We have so far considered the tau lepton and the corresponding neutrino, but only for definitiveness.
From the point of view of anomaly cancelation, which was our guiding principle in selecting the flavor symmetry, any lepton flavor could have been selected to be charged under this new symmetry. This means that it is possible to generate any flavor diagonal neutrino non-standard interaction ($\epsilon_{ee}$ or $\epsilon_{\mu\mu}$) in a similar way.  

Some of the constraints and future search strategies are different in this case and require a reanalysis. In particular, one important consequence of assigning the new $U(1)$ charges to the muon instead of tau would be the effects of \emph{non-universality} in bottom meson decays to electrons and muon. To this end, let us mention that our model is relevant to the recently reported anomaly \cite{LHCb} in $b\rightarrow s l^{+}l^{-}$ decays (specifically, the ratio $BR(B^{0}\rightarrow K^{\ast 0} \mu^{+} \mu^{-})/BR(B^{0}\rightarrow K^{\ast 0} e^{+} e^{-})$). This analysis will be reported elsewhere.

Another class of observables involves processes such as the $D-\bar D$ mixing, atomic parity violation, and M{\o}ller scattering: for a heavy mediator
they probe a combination of VEVs in the Higgs sector, related to the $X-Z$ mixing, while for a low mediator mass  they are also sensitive to the gauge coupling $g_{X}$. Although the current constraint from M{\o}ller scattering is not competitive with other bounds, the future MOLLER experiment at Jefferson Lab will improve the bound, and may even take a leading role in constraining some region of the parameter space.

A unique role is played by the process $\Upsilon \rightarrow \tau^{+}\tau^{-}$, which operates entirely in the third generation and gives the most direct access to the coupling $g_{X}$. Future $B$-factory data may improve the universality bound on $\Upsilon$ decays and further constrain third family gauge symmetries.

Finally, LHC searches for $X$ boson production may also play a role in constraining the model.
$Z$ decays to $\tau^+ \tau^- X$ and $b\bar bX$, although less constraining at the
present time, are also a direct probe of $g_X$. Thus, for $X$ masses below the $Z$ mass, dedicated searches taking into account the differences in
kinematics between two and three body decays of the $Z$ may be interesting venues for future exploration.
We also identify resonant $X$ boson production decaying to $\tau$ pairs in association with $b$-jets as a promising search for heavier masses. We thus  foresee the LHC phenomenology of our model to be rich.

Another area of phenomenology that deserves a close look is establishing astrophysical signatures of this scenario. The relevant discussion is deferred to a separate paper, mainly to keep the scope of the present work finite. In brief, supernova and stellar cooling considerations do yield constraints on certain parts of the parameter space (low mass and small coupling).

A number of theoretical directions must also be pursued. Among them is the origin of the neutrino mass in this framework. Phenomenologically, we know that the lepton mixing matrix is qualitatively different from the CKM one, and in particular third generation is in no way singled out in it. This perhaps suggests that the origins of the neutrino masses and quark masses are different. Indeed, in our model phenomenological viability \emph{requires} that neutrino masses be generated by certain operators suppressed by the scale of the new symmetry breaking. Importantly, this scale is expected to be at the TeV scale due to anomaly cancelation and therefore can be within reach of the LHC and future collider experiments and, we hope, will motivate future searches.

\section{Acknowledgements}

We thank KITP Santa Barbara for hospitality and support (National Science Foundation under Grant No. PHY11-25915) during the
``Present and Future Neutrino Physics'' workshop, where this work was started. PM thanks the Oklahoma State University and SLAC for kind hospitality during the completion of this manuscript.
We thank Lance Dixon, Sasha Khanov, Aneesh Manohar, Carlos Pe\~na, Michael Peskin, and Renata Zukanovich Funchal
for useful discussions.  This work is supported
by the U.S. Department of Energy Grants No. de-sc0010108 (KSB), DE-AC02-76SF00515 (AF) and de-sc0013699 (IM), and by the EU grants H2020-MSCA-ITN-2015/674896-Elusives (PM) and H2020-MSCA-2015-690575-InvisiblesPlus (PM). Fermilab is operated by the Fermi Research Alliance, LLC under contract No. DE-AC02-07CH11359 with the United States Department of Energy.

\begin{appendices}
\section{Contributions to $B_d$ and $B_s$ widths}\label{Appendix}
In the model proposed, $\alpha$ and $\beta$, defined in Eq.~(\ref{eq:mixingmatrix}), would
contribute to  $b\to X d$ and $b\to X s$ transitions. An alternative version of this model would have
the $U(1)_X$ charge of $\phi_1$ and $s$ changed to $-1/3$. In that case, which would interchange
the structure of the up and down Yukawa sectors in Eq.~(\ref{eq:interaction}), the off-diagonal
couplings of the down sector, which will be called $\alpha$ and $\beta$ as well, would also
contribute to $b$ FCNCs. Here we estimate the constraints on the off-diagonal couplings of the
down-type quarks (valid for both versions of the model). The $B_d-\bar B_d$ and $B_s-\bar B_s$
mixings constraints discussed in section~\ref{sec:model} would apply, though they are not very
stringent.

The process $B_d\to\pi
X$ would contribute to the total $B_d$ width by the following amount (assuming $M_X\ll m_{B_d}$)
\begin{equation}
  \Gamma(B_d\to\pi X) = \frac{1}{144\pi}|F_+(M_X^2)|^2g_X^2a^2\frac{m_{B_d}^3}{M_X^2},
\end{equation}
where $m_{B_d}$ is the $B_d$ mass, and $F_+(q^2)$ is a form factor which can be determined by
use of chiral perturbation theory for heavy hadrons (see e.g. Ref.~\cite{Burdman:2003rs}).
Requiring this partial width to be below the total width of the $B_d$ translates into the constraint
\begin{equation}
  g_X < 2.9\times 10^{-5} \left(\frac{M_X}{100~\MeV}\right)\left(\frac{|V_{ub}|}{a}\right).
\end{equation}
A similar process could be considered, with $B^+\to \pi^+ X$ and $X$ further decaying to $\mu^+\mu^-$, leading to $B^+\to \pi^+\mu^+\mu^-$. This branching ratio is measured to be $1.79\times 10^{-8}$~\cite{Agashe:2014kda} and would constrain (requiring the new contribution not to exceed the measurement)
\begin{equation}
  g_X<\frac{3.8\times 10^{-10}}{\sqrt{BR(X\to\mu^+\mu^-)}}
    \left(\frac{M_X}{100~\MeV}\right)\left(\frac{|V_{ub}|}{a}\right).
\end{equation}
For instance, for $a=V_{ub}$, if $t_\beta=20$ then the $\mu\mu$ branching ratio is $2.5\times10^{-7}$, and thus $g_X<8.4\times 10^{-7}(M_X/100~\MeV)$, which should be compared to best limit of the model in the text, arising from neutrino oscillations, $g_X<4\times10^{-3}(M_X/100~\MeV)$. For the $B_s$, the total width bound is stronger than the $B_d$ case by an order of magnitude, since the $Xbs$ coupling would be proportional to $V_{cb}$ which is ten times larger than $V_{ub}$. More precisely, $g_X < 2.8\times 10^{-6} (M_X/100~\MeV)$.

\end{appendices}


\begin{thebibliography}{99}

\bibitem{Peskin:1995ev}
  M.~E.~Peskin and D.~V.~Schroeder,
  Reading, USA: Addison-Wesley (1995) 842 p

\bibitem{Gross:1972pv}
  D.~J.~Gross and R.~Jackiw,
  Phys.\ Rev.\ D {\bf 6}, 477 (1972).
  doi:10.1103/PhysRevD.6.477



  \bibitem{Pati:1974yy}
   J.~C.~Pati and A.~Salam,
  Phys.\ Rev.\ D {\bf 10}, 275 (1974)
  Erratum: [Phys.\ Rev.\ D {\bf 11}, 703 (1975)].

  \bibitem{Marshak:1979fm}
  R.~E.~Marshak and R.~N.~Mohapatra,
  Phys.\ Lett.\ B {\bf 91}, 222 (1980).

  \bibitem{Wilczek:1979et}
  F.~Wilczek and A.~Zee,
  Phys.\ Lett.\  {\bf 88B}, 311 (1979).
  doi:10.1016/0370-2693(79)90475-1



\bibitem{Mohapatra:1980qe}
  R.~N.~Mohapatra and R.~E.~Marshak,
  Phys.\ Rev.\ Lett.\  {\bf 44}, 1316 (1980)
  Erratum: [Phys.\ Rev.\ Lett.\  {\bf 44}, 1643 (1980)].



\bibitem{Nelson:2007yq}
  A.~E.~Nelson and J.~Walsh,
  Phys.\ Rev.\ D {\bf 77}, 033001 (2008)
  doi:10.1103/PhysRevD.77.033001
  [arXiv:0711.1363 [hep-ph]].

\bibitem{Harnik:2012ni}
  R.~Harnik, J.~Kopp and P.~A.~N.~Machado,
  JCAP {\bf 1207}, 026 (2012)
  [arXiv:1202.6073 [hep-ph]].


\bibitem{holdom}
 B.~Holdom,
  Phys.\ Lett.\  {\bf 166B}, 196 (1986).

\bibitem{Wilczek:1978xi}
  F.~Wilczek and A.~Zee,
  Phys.\ Rev.\ Lett.\  {\bf 42}, 421 (1979).
  doi:10.1103/PhysRevLett.42.421

\bibitem{Appelquist} 
  T.~Appelquist and R.~Shrock,
  Phys.\ Lett.\ B {\bf 548}, 204 (2002)
  [hep-ph/0204141];
  Phys.\ Rev.\ Lett.\  {\bf 90}, 201801 (2003)
  [hep-ph/0301108].


\bibitem{He:1991qd}
  X.~G.~He, G.~C.~Joshi, H.~Lew and R.~R.~Volkas,
  Phys.\ Rev.\ D {\bf 44}, 2118 (1991).
  doi:10.1103/PhysRevD.44.2118


\bibitem{Baek:2001kca}
  S.~Baek, N.~G.~Deshpande, X.~G.~He and P.~Ko,
  Phys.\ Rev.\ D {\bf 64}, 055006 (2001)
  doi:10.1103/PhysRevD.64.055006
  [hep-ph/0104141].

\bibitem{Ma:2001md}
  E.~Ma, D.~P.~Roy and S.~Roy,
  Phys.\ Lett.\ B {\bf 525}, 101 (2002)
  doi:10.1016/S0370-2693(01)01428-9
  [hep-ph/0110146].

\bibitem{Salvioni:2009jp}
  E.~Salvioni, A.~Strumia, G.~Villadoro and F.~Zwirner,
  JHEP {\bf 1003}, 010 (2010)
  doi:10.1007/JHEP03(2010)010
  [arXiv:0911.1450 [hep-ph]].

\bibitem{Heeck:2011wj}
  J.~Heeck and W.~Rodejohann,
  Phys.\ Rev.\ D {\bf 84}, 075007 (2011)
  doi:10.1103/PhysRevD.84.075007
  [arXiv:1107.5238 [hep-ph]].

\bibitem{Harigaya:2013twa}
  K.~Harigaya, T.~Igari, M.~M.~Nojiri, M.~Takeuchi and K.~Tobe,
  JHEP {\bf 1403}, 105 (2014)
  doi:10.1007/JHEP03(2014)105
  [arXiv:1311.0870 [hep-ph]].

\bibitem{Carone:2013uh}
  C.~D.~Carone,
  Phys.\ Lett.\ B {\bf 721}, 118 (2013)
  doi:10.1016/j.physletb.2013.03.011
  [arXiv:1301.2027 [hep-ph]].

\bibitem{Altmannshofer:2014cfa}
  W.~Altmannshofer, S.~Gori, M.~Pospelov and I.~Yavin,
  Phys.\ Rev.\ D {\bf 89}, 095033 (2014)
  doi:10.1103/PhysRevD.89.095033
  [arXiv:1403.1269 [hep-ph]].

\bibitem{Farzan:2015doa}
  Y.~Farzan,
  Phys.\ Lett.\ B {\bf 748}, 311 (2015)
  doi:10.1016/j.physletb.2015.07.015
  [arXiv:1505.06906 [hep-ph]].

\bibitem{Farzan:2015hkd}
  Y.~Farzan and I.~M.~Shoemaker,
  JHEP {\bf 1607}, 033 (2016)
  doi:10.1007/JHEP07(2016)033
  [arXiv:1512.09147 [hep-ph]].

\bibitem{D'Ambrosio:2002ex}
  G.~D'Ambrosio, G.~F.~Giudice, G.~Isidori and A.~Strumia,
  Nucl.\ Phys.\ B {\bf 645}, 155 (2002)
  [hep-ph/0207036].

\bibitem{Wolfenstein:1977ue}
  L.~Wolfenstein,
  Phys.\ Rev.\  {\bf D17}, 2369-2374 (1978).

\bibitem{Valle:1987gv}
  J.~W.~F.~Valle,
  Phys.\ Lett.\ B {\bf 199}, 432 (1987).

 \bibitem{Roulet:1991sm}
  E.~Roulet,
  Phys.\ Rev.\ D {\bf 44}, R935 (1991).

\bibitem{Guzzo:1991hi}
  M.~M.~Guzzo, A.~Masiero and S.~T.~Petcov,
  Phys.\ Lett.\ B {\bf 260}, 154 (1991).

\bibitem{GonzalezGarcia:2001mp}
  M.~C.~Gonzalez-Garcia, Y.~Grossman, A.~Gusso and Y.~Nir,
  Phys.\ Rev.\ D {\bf 64}, 096006 (2001).

 \bibitem{Fornengo:2001pm}
  N.~Fornengo, M.~Maltoni, R.~Tomas and J.~W.~F.~Valle,
  Phys.\ Rev.\ D {\bf 65}, 013010 (2002).

 \bibitem{Davidson:2003ha}
  S.~Davidson, C.~Pena-Garay, N.~Rius and A.~Santamaria,
  JHEP {\bf 0303}, 011 (2003)

\bibitem{Friedland:2004pp}
  A.~Friedland, C.~Lunardini and C.~Pena-Garay,
  Phys.\ Lett.\ B {\bf 594}, 347 (2004).

 \bibitem{Friedland:2005vy}
  A.~Friedland and C.~Lunardini,
  Phys.\ Rev.\ D {\bf 72}, 053009 (2005).

 \bibitem{Antusch:2008tz}
  S.~Antusch, J.~P.~Baumann and E.~Fernandez-Martinez,
  Nucl.\ Phys.\ B {\bf 810}, 369 (2009).

\bibitem{Gavela:2008ra}
  M.~B.~Gavela, D.~Hernandez, T.~Ota and W.~Winter,
  Phys.\ Rev.\ D {\bf 79}, 013007 (2009).

\bibitem{GonzalezGarcia:2011my}
  M.~C.~Gonzalez-Garcia, M.~Maltoni and J.~Salvado,
  JHEP {\bf 1105}, 075 (2011).

\bibitem{Friedland:2011za}
  A.~Friedland, M.~L.~Graesser, I.~M.~Shoemaker and L.~Vecchi,
  Phys.\ Lett.\ B {\bf 714}, 267 (2012)

\bibitem{Friedland:2012tq}
  A.~Friedland and I.~M.~Shoemaker,
  arXiv:1207.6642 [hep-ph].

\bibitem{Gonzalez-Garcia:2015qrr}
  M.~C.~Gonzalez-Garcia, M.~Maltoni and T.~Schwetz,
  Nucl.\ Phys.\ B {\bf 908}, 199 (2016)

\bibitem{Agashe:2014kda}
  K.~A.~Olive {\it et al.} [Particle Data Group Collaboration],
  Chin.\ Phys.\ C {\bf 38}, 090001 (2014).


\bibitem{vissani}
F.~Vissani,
  Phys.\ Rev.\ D {\bf 57}, 7027 (1998)
  doi:10.1103/PhysRevD.57.7027
  [hep-ph/9709409].


\bibitem{Baumgart:2009tn}
  M.~Baumgart, C.~Cheung, J.~T.~Ruderman, L.~T.~Wang and I.~Yavin,
  JHEP {\bf 0904}, 014 (2009)
  [arXiv:0901.0283 [hep-ph]].


\bibitem{Branco:2011iw}
  G.~C.~Branco, P.~M.~Ferreira, L.~Lavoura, M.~N.~Rebelo, M.~Sher and J.~P.~Silva,
  Phys.\ Rept.\  {\bf 516}, 1 (2012)
  [arXiv:1106.0034 [hep-ph]].

\bibitem{Khachatryan:2016vau}
  G.~Aad {\it et al.} [ATLAS and CMS Collaborations],
  JHEP {\bf 1608}, 045 (2016)
  [arXiv:1606.02266 [hep-ex]].


\bibitem{CMS:2013ada}
  CMS Collaboration [CMS Collaboration],
  CMS-PAS-HIG-13-014.


\bibitem{Aad:2015kna}
  G.~Aad {\it et al.} [ATLAS Collaboration],
  Eur.\ Phys.\ J.\ C {\bf 76}, no. 1, 45 (2016)
  [arXiv:1507.05930 [hep-ex]].


\bibitem{Hermann:2012fc}
  T.~Hermann, M.~Misiak and M.~Steinhauser,
  JHEP {\bf 1211}, 036 (2012)
  [arXiv:1208.2788 [hep-ph]].


\bibitem{Aad:2015typ}
  G.~Aad {\it et al.} [ATLAS Collaboration],
  JHEP {\bf 1603}, 127 (2016)
  [arXiv:1512.03704 [hep-ex]].


\bibitem{Essig:2013lka}
  R.~Essig {\it et al.},
  arXiv:1311.0029 [hep-ph].

\bibitem{Whally:2003}
  M.~R.~Whalley,
  J. Phys. G: Nuclear and Particle Physics {\bf 29} (2003), no 12A A1.

\bibitem{hepdata}
http://hepdata.cedar.ac.uk/review/rsig/


\bibitem{delAmoSanchez:2010bt}
  P.~del Amo Sanchez {\it et al.} [BaBar Collaboration],
  Phys.\ Rev.\ Lett.\  {\bf 104}, 191801 (2010)
  [arXiv:1002.4358 [hep-ex]].


\bibitem{Manohar:2003xv}
  A.~V.~Manohar and P.~Ruiz-Femenia,
  Phys.\ Rev.\ D {\bf 69}, 053003 (2004)
  [hep-ph/0311002].


\bibitem{Blum:2009sk}
  K.~Blum, Y.~Grossman, Y.~Nir and G.~Perez,
  Phys.\ Rev.\ Lett.\  {\bf 102} (2009) 211802
  [arXiv:0903.2118 [hep-ph]].

\bibitem{Babu:2009nn}
  K.~S.~Babu and Y.~Meng,
  Phys.\ Rev.\ D {\bf 80}, 075003 (2009)
  [arXiv:0907.4231 [hep-ph]].


\bibitem{Babu:1999me}
  K.~S.~Babu and S.~Nandi,
  Phys.\ Rev.\ D {\bf 62}, 033002 (2000)
  [hep-ph/9907213].

\bibitem{Golowich:2009ii} 
  E.~Golowich, J.~Hewett, S.~Pakvasa and A.~A.~Petrov,
  Phys.\ Rev.\ D {\bf 79}, 114030 (2009)
  [arXiv:0903.2830 [hep-ph]].


\bibitem{Burdman:2003rs}
  G.~Burdman and I.~Shipsey,
  Ann.\ Rev.\ Nucl.\ Part.\ Sci.\  {\bf 53}, 431 (2003)
  [hep-ph/0310076].

\bibitem{Babu:1987xe}
  K.~S.~Babu, X.~G.~He, X.~Li and S.~Pakvasa,
  Phys.\ Lett.\ B {\bf 205}, 540 (1988).

\bibitem{Marciano:1996wy} 
  W.~J.~Marciano and Z.~Parsa,
  Phys.\ Rev.\ D {\bf 53}, no. 1, R1 (1996).


\bibitem{Anisimovsky:2004hr}
  V.~V.~Anisimovsky {\it et al.} [E949 Collaboration],
  Phys.\ Rev.\ Lett.\  {\bf 93}, 031801 (2004)
  [hep-ex/0403036].

\bibitem{Ball:2004ye} 
  P.~Ball and R.~Zwicky,
  Phys.\ Rev.\ D {\bf 71}, 014015 (2005)
  [hep-ph/0406232].


\bibitem{Artamonov:2008qb}
  A.~V.~Artamonov {\it et al.} [E949 Collaboration],
  Phys.\ Rev.\ Lett.\  {\bf 101}, 191802 (2008)
  [arXiv:0808.2459 [hep-ex]].

\bibitem{Inami:1980fz}
  T.~Inami and C.~S.~Lim,
  Prog.\ Theor.\ Phys.\  {\bf 65}, 297 (1981)
  Erratum: [Prog.\ Theor.\ Phys.\  {\bf 65}, 1772 (1981)].

\bibitem{Hall:1981bc}
  L.~J.~Hall and M.~B.~Wise,
  Nucl.\ Phys.\ B {\bf 187}, 397 (1981).

\bibitem{Frere:1981cc}
  J.~M.~Frere, J.~A.~M.~Vermaseren and M.~B.~Gavela,
  Phys.\ Lett.\  {\bf 103B}, 129 (1981).

\bibitem{Freytsis:2009ct}
  M.~Freytsis, Z.~Ligeti and J.~Thaler,
  Phys.\ Rev.\ D {\bf 81}, 034001 (2010)
  [arXiv:0911.5355 [hep-ph]].

\bibitem{Davoudiasl:2012ag}
  H.~Davoudiasl, H.~S.~Lee and W.~J.~Marciano,
  Phys.\ Rev.\ D {\bf 85}, 115019 (2012)
  [arXiv:1203.2947 [hep-ph]].


\bibitem{Gonzalez-Garcia:2013usa}
  M.~C.~Gonzalez-Garcia and M.~Maltoni,
  JHEP {\bf 1309}, 152 (2013)
  [arXiv:1307.3092].



\bibitem{Porsev:2010de}
  S.~G.~Porsev, K.~Beloy and A.~Derevianko,
  Phys.\ Rev.\ D {\bf 82}, 036008 (2010)
  [arXiv:1006.4193 [hep-ph]].

\bibitem{Aad:2015txa}
  G.~Aad {\it et al.} [ATLAS Collaboration],
  JHEP {\bf 1601}, 172 (2016)
  [arXiv:1508.07869 [hep-ex]].

\bibitem{Anthony:2003ub}
  P.~L.~Anthony {\it et al.} [SLAC E158 Collaboration],
  Phys.\ Rev.\ Lett.\  {\bf 92}, 181602 (2004)
  [hep-ex/0312035].

\bibitem{Marciano:1979cd}
  W.~J.~Marciano and D.~Wyler,
  Z.\ Phys.\ C {\bf 3}, 181 (1979).


    \bibitem{Chatrchyan:2012sv}
  S.~Chatrchyan {\it et al.} [CMS Collaboration],
  Phys.\ Rev.\ Lett.\  {\bf 110}, no. 8, 081801 (2013)
  [arXiv:1210.5629 [hep-ex]].

\bibitem{ATLAS:2013oea}
  G.~Aad {\it et al.} [ATLAS Collaboration],
  JHEP {\bf 1306}, 033 (2013)
  [arXiv:1303.0526 [hep-ex]].


\bibitem{Nir:1990yq}
  Y.~Nir and D.~J.~Silverman,
  Phys.\ Rev.\ D {\bf 42}, 1477 (1990).


\bibitem{Buras:1990fn}
  A.~J.~Buras, M.~Jamin and P.~H.~Weisz,
  Nucl.\ Phys.\ B {\bf 347}, 491 (1990).


\bibitem{Lenz:2012mb}
  A.~Lenz,
  arXiv:1205.1444 [hep-ph].


\bibitem{Bjorken:2009mm}
  J.~D.~Bjorken, R.~Essig, P.~Schuster and N.~Toro,
  Phys.\ Rev.\ D {\bf 80}, 075018 (2009)
  [arXiv:0906.0580 [hep-ph]].


\bibitem{Studenikin:1998cs}
  A.~I.~Studenikin,
  hep-ph/9808219.


\bibitem{Jegerlehner:2009ry}
  F.~Jegerlehner and A.~Nyffeler,
  Phys.\ Rept.\  {\bf 477}, 1 (2009)
  [arXiv:0902.3360 [hep-ph]].


\bibitem{Ablikim:2014uzh}
  M.~Ablikim {\it et al.} [BESIII Collaboration],
  Phys.\ Rev.\ D {\bf 90}, no. 1, 012001 (2014)
  [arXiv:1405.1076 [hep-ex]].


\bibitem{Bellini:2011rx}
  G.~Bellini {\it et al.},
  Phys.\ Rev.\ Lett.\  {\bf 107}, 141302 (2011)
  [arXiv:1104.1816 [hep-ex]].

\bibitem{Beda:2009kx}
  A.~G.~Beda, E.~V.~Demidova, A.~S.~Starostin, V.~B.~Brudanin, V.~G.~Egorov, D.~V.~Medvedev, M.~V.~Shirchenko and T.~Vylov,
  Phys.\ Part.\ Nucl.\ Lett.\  {\bf 7}, 406 (2010)
  [arXiv:0906.1926 [hep-ex]].


\bibitem{Vilain:1994qy}
  P.~Vilain {\it et al.} [CHARM-II Collaboration],
  Phys.\ Lett.\ B {\bf 335}, 246 (1994).


\bibitem{Davidson:2003ha}
  S.~Davidson, C.~Pena-Garay, N.~Rius and A.~Santamaria,
  JHEP {\bf 0303}, 011 (2003)
  [hep-ph/0302093].


\bibitem{Deniz:2009mu}
  M.~Deniz {\it et al.} [TEXONO Collaboration],
  Phys.\ Rev.\ D {\bf 81}, 072001 (2010)
  [arXiv:0911.1597 [hep-ex]].


\bibitem{AguilarArevalo:2007it}
  A.~A.~Aguilar-Arevalo {\it et al.} [MiniBooNE Collaboration],
  Phys.\ Rev.\ Lett.\  {\bf 98}, 231801 (2007)
  [arXiv:0704.1500 [hep-ex]].


\bibitem{Aguilar-Arevalo:2013pmq}
  A.~A.~Aguilar-Arevalo {\it et al.} [MiniBooNE Collaboration],
  Phys.\ Rev.\ Lett.\  {\bf 110}, 161801 (2013)
  [arXiv:1207.4809 [hep-ex], arXiv:1303.2588 [hep-ex]].


\bibitem{Auerbach:2001wg}
  L.~B.~Auerbach {\it et al.} [LSND Collaboration],
  Phys.\ Rev.\ D {\bf 63}, 112001 (2001)
  [hep-ex/0101039].


\bibitem{Zeller:2001hh}
  G.~P.~Zeller {\it et al.} [NuTeV Collaboration],
  Phys.\ Rev.\ Lett.\  {\bf 88}, 091802 (2002)
  Erratum: [Phys.\ Rev.\ Lett.\  {\bf 90}, 239902 (2003)]
  [hep-ex/0110059].


\bibitem{Acciarri:2000uh}
  M.~Acciarri {\it et al.} [L3 Collaboration],
  Phys.\ Lett.\ B {\bf 489}, 81 (2000)
  [hep-ex/0005028].

\bibitem{Batell:2009di}
  B.~Batell, M.~Pospelov and A.~Ritz,
  Phys.\ Rev.\ D {\bf 80}, 095024 (2009)
  [arXiv:0906.5614 [hep-ph]].


\bibitem{Wong:2006nx}
  H.~T.~Wong {\it et al.} [TEXONO Collaboration],
  Phys.\ Rev.\ D {\bf 75}, 012001 (2007)
  [hep-ex/0605006].


\bibitem{Chen:2014dsa}
  J.~W.~Chen, H.~C.~Chi, H.~B.~Li, C.-P.~Liu, L.~Singh, H.~T.~Wong, C.~L.~Wu and C.~P.~Wu,
  Phys.\ Rev.\ D {\bf 90}, no. 1, 011301 (2014)
  [arXiv:1405.7168 [hep-ph]].

\bibitem{LHCb}
Simone Bifani [LHCb Collaboration],  
``Search for New Physics with $b\rightarrow sll$ decays  @ LHCb'', CERN seminar, April 18 (2017).

\end{thebibliography}
\end{document}